\documentclass[letterpaper,twocolumn,10pt]{article}
\usepackage{usenix}
\usepackage{tikz}
\usepackage{amsmath}
\usepackage{graphicx}
\usepackage{alltt}
\usepackage{ulem}
\usepackage{array}
\usepackage{booktabs}
\usepackage{multirow}

\usepackage{tabularx}
\usepackage{hyperref}
\usepackage{titlesec}
\usepackage[hang,flushmargin]{footmisc}
\usepackage{titling}
\begin{document}

\date{}

\title{\Large \bf Generated Data with Fake Privacy: \\Hidden Dangers of Fine-tuning Large Language Models on Generated Data}


\author{
{\rm Atilla Akkus$^{*}$}\\
Bilkent University
\and
{\rm Masoud Poorghaffar Aghdam$^{*}$}\\
Bilkent University
\and
{\rm Mingjie Li$^*$}\\
CISPA
\and
{\rm Junjie Chu}\\
CISPA
\and
{\rm Michael Backes}\\
CISPA 
\and
{\rm Yang Zhang}\\
CISPA
\and
{\rm Sinem Sav}\\
Bilkent University
}

\maketitle
\def\thefootnote{*}\footnotetext{These authors contributed equally to this work}

\begin{abstract}
Large language models (LLMs) have demonstrated significant success in various domain-specific tasks, with their performance often improving substantially after fine-tuning. However, fine-tuning with real-world data introduces privacy risks. To mitigate these risks, developers increasingly rely on synthetic data generation as an alternative to using real data, as data generated by traditional models is believed to be different from real-world data. However, with the advanced capabilities of LLMs, the distinction between real data and data generated by these models has become nearly indistinguishable. This convergence introduces similar privacy risks for generated data to those associated with real data. In this paper, we present an empirical analysis of this underexplored issue by investigating a key question: "\textit{Does fine-tuning with LLM-generated data enhance privacy, or does it pose additional privacy risks?}" Our study investigates this question by examining the structural characteristics of data generated by LLMs, focusing on two primary fine-tuning approaches: supervised fine-tuning (SFT) with unstructured (plain-text) generated data and self-instruct tuning. In the scenario of SFT, the data is put into a particular instruction tuning format used by previous studies. We use Personal Information Identifier (PII) leakage and Membership Inference Attacks (MIAs) on the Pythia Model Suite and Open Pre-trained Transformer (OPT) to measure privacy risks. Notably, after fine-tuning with unstructured generated data, the rate of successful PII extractions for Pythia increased by over $20\%$, highlighting the potential privacy implications of such approaches. Furthermore, the ROC-AUC score of MIAs for Pythia-6.9b, the second biggest model of the suite, increases over $40\%$  after self-instruct tuning. 
Our results indicate the potential privacy risks associated with fine-tuning LLMs using generated data, underscoring the need for careful consideration of privacy safeguards in such approaches.

\end{abstract}

\section{Introduction}
Recently, large language models (LLMs) such as GPT-4~\cite{gpt4_blog}, LLaMA-3~\cite{llama3_license}, and Mistral~\cite{mistral_license} have demonstrated considerable success in text generation and have been extensively deployed for a variety of specific tasks, particularly as customized chatbots. 
The impressive capabilities of these LLMs are largely attributed to the vast pre-training datasets sourced from the Internet or data providers ~\cite{BMRSKDNSSAAHKHCRZWWHCSLGCCBMRSA20}. The choice of training data plays a critical role in the performance of LLMs. As a result, many LLM providers, such as OpenAI and Meta, opt to keep their training data selection confidential. However, the training data often contains privacy-sensitive data from real individuals~\cite{bommasani2022opportunitiesrisksfoundationmodels, carlini2021extracting}. To assess the potential privacy risks on sensitive information or private training data, researchers have proposed numerous well-designed attacks associated with LLMs, such as membership inference attacks (MIA)\cite{SIHS21, yeom2018privacyriskmachinelearninglossbasedattack, shi2024detectingpretrainingdatalargeminkattack,carlini2021extracting,carlini2022membershipinferenceattacksprinciplesreferenceattack}, Personally Identifiable Information (PII) attacks~\cite{LSSTWB23}, and data extraction attacks~\cite{nasr2023scalable,shi2024detectingpretrainingdatalargeminkattack,carlini2021extracting}. 

However, recent research~\cite{JanusInterface} highlights that fine-tuning an LLM on datasets overlapping with its pre-training data can pose privacy risks, especially to closely related pre-training portions. In~\cite{JanusInterface}, researchers fine-tuned LLMs using a small subset of the pre-training data and observed that this process also enhances the model's memorization of other data points related to the fine-tuning data. The fine-tuning process can amplify an LLM's memorization capabilities, potentially leading to privacy risks, such as the extraction of sensitive information~\cite{LSSTWB23}. Consequently, fine-tuning with real datasets raises significant privacy concerns, especially when these datasets overlap with the model's pre-training data. However, no previous research has explored the implications of fine-tuning with \textit{generated} data.

Similar to traditional machine learning, LLMs can also leverage generated data for fine-tuning. Notable examples include Alpaca~\cite{alpaca} for instruction tuning and HH-RLHF~\cite{BJNACDDFGHJKKCEEHHHJKLNOABCMOMK22} for preference optimization, among others. Moreover, researchers have developed various prompting techniques to help LLMs generate high-quality data for fine-tuning. For example, developers can use concise, human-written prompts to guide LLMs in generating content for fine-tuning. Alternatively, they can provide input-output pairs for specific seed tasks, which serve as prompts for the model to generate additional task-specific examples with corresponding pairs, facilitating further fine-tuning. These generated datasets greatly improve the performance of LLMs and are widely adopted due to their flexibility and low costs. This naturally raises the question: \textit{Does fine-tuning on entirely synthetic datasets generated by LLMs introduce privacy risks?} This inquiry has not been addressed by previous research in terms of concrete privacy risks. Our findings suggest that despite common belief, generated data does not mitigate but exacerbates the risks.

\subsection{Threat Model}

We explore scenarios in which LLM developers initially train a model using their proprietary datasets and subsequently fine-tune it to perform various domain-specific tasks before making it publicly available. Recognizing the risk of private data leakage that may arise from fine-tuning specialized LLMs with portions of the original training set, developers opt to use generated data for this process. Our paper evaluates the potential privacy risks of fine-tuning using generated data. Since most LLM developers offer access to their fine-tuned models solely through a query-based API, potential attackers would be limited to querying the models to extract sensitive information. However, we also consider the more severe case in which attackers can access the returned logits of the outputs. To assess privacy risks, we employ PII extraction and score-based MIA techniques.

\subsection{Our Work} 

To study the aforementioned risks, we begin by experimenting with fine-tuning LLMs using various types of generated data. We then employ MIA and PII attacks to evaluate the privacy risks. Our study primarily examines the two most common fine-tuning scenarios for language models: supervised fine-tuning (SFT) with unstructured data and instruction tuning.

The SFT with unstructured data scenario is designed to enhance the model's performance across various domains, such as improving comprehension or reasoning on emails. In this case, we prompt LLMs with email-specific prefixes and use their completions for fine-tuning. This is explained in more detail in the Section \ref{preliminaryfinetuning}.

On the other hand, the self-instruct approach feeds existing tasks and input-output pairs into a capable LLM -such as GPT 3.5 as used by \cite{wang2023selfinstructaligninglanguagemodels}- to generate similar tasks with new input-output pairs. Fine-tuning with this generated data not only enhances domain-specific capabilities but also improves the model's ability to better follow user prompts. The details are listed as follows.


\par\noindent\textbf{Risks on Fine-Tuning with Unstructured Generated Data.} Following the setting in former work~\cite{JanusInterface}, we use the Enron email dataset to evaluate the potential privacy risks on fine-tuned Pythia models. We first use Pythia-12b to generate an email dataset, and then fine-tune the Pythia model with different model sizes on these generated datasets. Then, we conduct PII attacks following Wang et al.'s~\cite{wang2024decodingtrustcomprehensiveassessmenttrustworthiness} setting on both the pre-trained model and fine-tuned models. The results demonstrate that supervised fine-tuning amplifies privacy risks even in unstructured generated data (Section~\ref{sec-3-2}). After that, we ran experiments to analyze the privacy risks and found that the template and quality of generated data are the main factors that may influence PII's success rate (Section~\ref{sec-3-3}). 

In addition to these experiments, we conduct further evaluations using Facebook's OPT model, which has no overlap with Enron or any email data in its training set. For this evaluation, we modify the Enron dataset to minimize overlap with the sensitive data in Pythia's training data. These experiments not only extend the scope of our analysis but also serve to verify the results observed in our previous experiments, confirming the trends and insights derived from the Pythia-based evaluations in more challenging scenarios.

\par\noindent\textbf{Risks on Fine-Tuning with Self-Instruct.} In Section \ref{Self-Instruct} we conduct experiments following the "self-instruct" tuning pipeline, as illustrated in previous research~\cite{wang2023selfinstructaligninglanguagemodels}. Our aim is to examine the potential privacy risks associated with Pythia's pre-trained datasets, The Pile~\cite{gao2020pile800gbdatasetdiverse}. We choose the FreeLaw~\cite{gao2020pile800gbdatasetdiverse} subset of The Pile for the privacy-sensitive nature of the law domain. In line with the self-instruct procedure, we initially designed 64 task descriptions focusing on legal expertise and 75 related input-output pairs (denoted as seed tasks) based on the FreeLaw dataset. After that, we prompt LLama-3 with these seed tasks to procure the generated data for fine-tuning, including task descriptions, related information, and answers. 

Fine-tuning Pythia models with the generated data, we can obtain LLMs that exhibit enhanced performance on legal question-answering tasks. Then we conduct the score-based MIA method following Duan et al.~\cite{duan2024membershipinferenceattackswork}'s setting, on the self-instruct tuned Pythia models and their pre-trained version. The results reveal that the AUC ROC score of MIA on FreeLaw datasets enjoys nearly $20\%$ improvement compared to the pre-trained model. These findings highlight that leveraging self-instructed data generated by LLMs can intensify the model's susceptibility to privacy vulnerabilities. Further investigation reveals that the primary factor influencing the models' privacy is the quality of the generated data. We summarize our contributions as follows:
\begin{itemize}
\item We evaluate the privacy risks of supervised fine-tuning in LLMs using generated data without an instructional structure, specifically through a PII attack. The results demonstrate that fine-tuning with generated email data increases the success rate of PII attacks by over $50\%$ compared to the pre-trained model. This suggests that training on generated raw data within the same domain can significantly amplify the privacy leakage associated with the LLM's pre-training datasets.

\item We evaluate the privacy risks associated with fine-tuned LLMs using instruction-based data. Our analysis shows that self-instruct tuning on law-related tasks increases the model's vulnerability within the law-related subset from its pre-training data. Specifically, the AUC-ROC score for a reference-based MIA attack on the fine-tuned Pythia-6.9b model increased by $20\%$ compared to the pre-trained model. These results suggest that self-instruct tuning can exacerbate privacy risks, especially in domains closely related to self-instruct tasks.

\item We further investigate the causes of such a phenomenon and find that the heightened privacy risk stems from the high quality of the generated data and its similarity to the pre-training datasets. Additionally, we explore the key factors contributing to these potential privacy risks and propose several practical methods to mitigate them.
\end{itemize}

\section{Preliminaries}
In this section, we summarize LLM's prompting, pre-training, and various fine-tuning methods.
\vspace{-0.5em}
\subsection{LLM Prompting}
Prompting methods are strategies used to obtain specific responses from LLMs by designing the input text in particular ways. These methods can be used to increase the comprehension of LLMs on the task, often with purposes such as enhancing response quality or format. Some prompting methods depend on LLMs ability to recognize patterns and generate the response based on them. One such effective technique is \textit{few-shot prompting}~\cite{BMRSKDNSSAAHKHCRZWWHCSLGCCBMRSA20}, where the prompt includes a handful of example input-output pairs that demonstrate the desired task. By presenting these examples, the LLM can infer the underlying task structure and produce appropriate responses to new inputs. For instance, to translate a particular text from English to German, the prompt might include "Hello" → "Hallo" and "Good morning" → "Guten Morgen," enabling the LLM to translate "Thank you" → "Danke."

\subsection{LLM's Pre-training}

Pretraining large language models (LLMs) has significantly advanced with the development of transformer-based architectures, compared with former approaches like Word2Vec and GloVe. Notably, GPT-based models \cite{radford2018improving} pioneered the autoregressive pretraining paradigm, where the model learns to predict the next token in a sequence. GPT-2 \cite{radford2019language} further demonstrated the capabilities of large-scale unsupervised learning, setting new benchmarks in various NLP tasks. Building on these successes, GPT-3 \cite{BMRSKDNSSAAHKHCRZWWHCSLGCCBMRSA20} introduced even larger models, with 175 billion parameters, and showcased remarkable performance across a wide range of tasks without requiring task-specific fine-tuning. More recently, open-sourced models such as LLaMA \cite{touvron2023llama} have emerged, aiming to provide highly efficient alternatives by optimizing training and scaling strategies. LLaMA models, like GPT, are designed to excel in language understanding and generation while being more accessible for research and applications. Additionally, models like Pythia \cite{biderman2023pythiasuiteanalyzinglarge} and Mistral \cite{mistral_license} have contributed to making large-scale autoregressive models available to a broader community, encouraging further exploration and refinement of pretraining techniques. Despite these advancements, challenges related to model bias, computational cost, and interpretability remain central to ongoing research in the field of LLM pretraining.

\subsection{Fine-tuning Methods}\label{preliminaryfinetuning}
\par\noindent\textbf{Fine-Tuning with Unstructured Raw Texts.} To enhance the capabilities of LLMs across various domain-specific tasks, users can fine-tune them further using specialized datasets, such as those from biomedicine, law, and finance domains. The fine-tuning process resembles the pre-training of LLMs but typically involves a smaller dataset. Because pre-trained LLMs already have a solid understanding of language intrinsics, fine-tuning them on domain-specific datasets can yield competitive performances. Additionally, several closed-source LLM providers, such as OpenAI, offer APIs for fine-tuning using unstructured raw text, enabling users to further optimize model performance for their specific needs.

\par\noindent\textbf{Instruction Tuning and Self-Instruct.} Instruction tuning is a popular technique employed to enhance the ability of large language models (LLMs) to follow user prompts, thereby producing more accurate responses. In contrast to traditional fine-tuning, which utilizes raw textual data, instruction tuning necessitates the use of manually crafted instructions, user-generated prompts (inputs), and expected answers (outputs). Consequently, the collection of such data is heavily dependent on manual labeling, which is often resource-intensive. To address the challenges associated with gathering instruction data, researchers have proposed the "self-instruct" method. This approach involves using advanced LLMs to generate instruction-tuning samples, which can then be utilized for fine-tuning purposes.

To generate instruct data with good quality, Self-Instruct first takes an initial dataset of instructions and their corresponding input-output examples, termed 'seed tasks'. 
For example, an instruction might be, "What is the name of the victim in the following legal document?" The corresponding input-output examples would consist of legal documents as inputs and the identified victims mentioned in them as outputs. The quality and diversity of seed tasks are vital for the efficacy of the procedure. Once the seed tasks are ready, the rest of the procedure depends on the \textit{Generator}, and the \textit{Target model} (see below), which are not necessarily distinct.
\begin{itemize}
    \item Bootstrapping Tasks. Depending on the seed tasks, new tasks are generated by the generator. The accurate and creative generation of these tasks is achieved by few-shot prompting with seed tasks.
    \item Bootstrapping Examples. For each generated task, the generator creates new input-output examples using a similar few-shot prompting approach, incorporating instruction-input-output triples. Generated examples not in the desired form are excluded from the next step.
    \item Training the Target Model. The generated tasks and their examples are combined and formatted inside an instruction template to train the target.
\end{itemize}
\par\noindent\textbf{Parameter-efficient fine-tuning (PEFT) methods.} We describe below the PEFT methods used in this study:
\par\noindent\textbf{LoRA and Quantization}
Hu et al. propose the Low Rank Adaptation (LoRA) ~\cite{hu2021loralowrankadaptationlarge} for efficient fine-tuning of models without utility loss. LoRA reduces occupied memory during fine-tuning by "freezing" a large portion of model parameters and updating the trainable parameters with low-rank approximation (i.e., adapter) of the update matrix. The \textit{low-rank approximation} involves decomposing a high-dimensional matrix into the product of two lower-dimensional matrices, reducing computational complexity. The \textit{update matrix} refers to the changes applied to the original model parameters in each step. The adapter is optimized with respect to the loss function and multiplied by the scale factor to control the magnitude of the updates. This approach enables the integration of the base model with various adapters, which are significantly smaller in size compared to fully fine-tuned models. Besides the LoRA method, various methods are proposed to reduce LoRA's parameter \cite{kopiczko2023vera}, increase safety \cite{li2025salora} and etc \cite{jiang2024mora}.

To further reduce the computational cost of fine-tuning large models, Dettmers et al. introduced Quantized Low Rank Adaptation (QLoRA) method ~\cite{dettmers2023qloraefficientfinetuningquantized}. QLoRA uses block-wise quantization which divides the model parameters into smaller blocks and quantizes each block separately. This technique minimizes precision loss and reduces computational overhead, enabling the training of the quantized LoRA adapter and we use QLoRA for Pythia experiments.

\par\noindent\textbf{DoRA}, an improvement over LoRA, was introduced in~\cite{liu2024dora}. Unlike LoRA, which primarily focuses on low-rank adaptation, DoRA incorporates an additional trainable parameter: magnitude. This parameter enhances the model's flexibility, enabling faster convergence while achieving higher precision during fine-tuning. We use DoRA to fine-tune the Facebook OPT 1.4b parameter model and the Pythia 2.8b parameter model on datasets that do not overlap with their training data.



\subsection{Models}
 We choose Pythia and Open Pre-trained Transformer (OPT) language models as the target models to evaluate the potential privacy risks. We also use the powerful Llama-3 as the self-instruct method's generator to generate high-quality fine-tuning data. Details for these models are listed as follows.
\par\noindent\textbf{Pythia Suite} is developed by EleutherAI~\cite{biderman2023pythiasuiteanalyzinglarge} and provides open-source LLMs of varying sizes. The models at each size are trained on both the standard and deduplicated version of The Pile~\cite{gao2020pile800gbdatasetdiverse}. We use Pythia models with parameter sizes of 410m, 1.4b, 2.8b, and 6.9b as target models, while the 12b model serves as the generator in Section~\ref{Enron}.


\par\noindent\textbf{Llama-3-8b-Instruct} is the smallest model in Meta's open-source Llama-3 collection~\cite{MetaLlama3}. This model is chosen for its strong instruction-following capabilities and relatively compact size. It is used for creative generation tasks in Section~\ref{Self-Instruct}.

\par\noindent\textbf{OPT Language Models} were introduced by Meta AI to provide researchers with access to high-performance language models~\cite{zhang2022optopenpretrainedtransformerFacebookOPT}. These models are designed to approximately match the size and performance of the GPT-3 family of models. Pre-trained versions of OPT are available in various sizes, ranging from 125m to 66b parameters. In this work, we employ the 1.3 billion and 2.7 billion parameter versions of the OPT model. This choice was made because a subset of the Pile dataset ~\cite{gao2020pile800gbdatasetdiverse}, including CommonCrawl, DM Mathematics, Project Gutenberg, HackerNews, OpenSubtitles, OpenWebText2, USPTO, and Wikipedia, was used in the training of these models. Notably, the Enron subset is excluded from this dataset. Consequently, the training data for the OPT models does not overlap with the data used in our method, ensuring data independence and mitigating potential biases.

\subsection{Datasets}
\label{preliminaryDatasets}

We evaluate the privacy risks on Pythia's training dataset, the Pile. Especially, we use its Enron subset for plain-text fine-tuning and FreeLaw subset for instruction-tuning. 
\par\noindent\textbf{Pile:} The Pile~\cite{gao2020pile800gbdatasetdiverse} involves 800GB data from various sources including Internet forums, video subtitles, and academic texts. The Pile has been used for various model's pre-training such as GPT-Neo and Pythia. It consists of 22 smaller datasets including Enron and Freelaw corpora.

\par\noindent\textbf{Enron:}\label{enronDataset} Enron corpus~\cite{theenroncorpus} is a Pile subset containing different email conversations. We use a preprocessed version of the dataset shared by~\cite{wang2024decodingtrustcomprehensiveassessmenttrustworthiness} which consists of 3330 samples. For each sample, the original sample in Enron is split into the following columns:
\begin{enumerate}
\item \textbf{Prompt.} First part of each selected conversation. Used to prompt the LLM to generate the continuation.
\item \textbf{Continuation.} The second part of each selected conversation completes the logical flow introduced by the prompt. LLM's generation is compared with this column in terms of language, semantic similarity, and coherence.
\item \textbf{Name.} The name of the target person that is mentioned in the conversation.
\item \textbf{Email.} The email of the target person. This is not given in the conversation but has been asked the model to generate based on the owner's name or the context introduced in the correspondence. For instance, the model may be requested to generate the email address of a person named John Doe and is told to be working at \textit{Lipsum Energy Inc.}, which may be john.doe@lipsumenergy.com.
\end{enumerate}
\textbf{Psedonymized Enron:} We create an extended version of the Enron dataset, referred to as the Pseudonymized Enron dataset, for experimental purposes where the pretraining data does not overlap with the generated fine-tuning data. In this dataset, the original names and email addresses from Enron were replaced with synthetic data generated using the Faker library~\cite{faker}. Note that due to the randomness and the limited selection of names and emails in this library, some pseudonymous names and email addresses might appear multiple times. Names are identified and replaced using regular expressions to match capitalized words, while email addresses are detected by identifying patterns containing a domain following the "@" symbol.
\par\noindent\textbf{FreeLaw: }
FreeLaw is an open-source dataset related to the legal domain. It is a subset of The Pile that is obtained from the CourtListener~\cite{CourtListener} project. CourtListener includes a large number of legal opinions from federal
and state courts. It consists of numerous modalities of legal proceedings, including dockets, bibliographic information on judges, etc. Following Pile's setting, we only focus on court opinions due to an abundance of full-text entries. 

\section{Privacy Risks on Fine-Tuning with Unstructured Generated Data}\label{Enron}

In this section, we explore the potential risks of supervised fine-tuning with unstructured generated data. Similar to the scenario presented by Chen et al.~\cite{JanusInterface}, we assume that model owners seek to improve their model's performance in the email domain through fine-tuning. However, we introduce an additional strict assumption: the model owners lack access to real fine-tuning data. Thus, they can only rely on other LLMs to generate email-related data for this fine-tuning process. After the fine-tuning, we perform the PII extraction attack on the Enron dataset to evaluate the potential privacy risks of the fine-tuned models.
\vspace{-0.5em}
\subsection{Experimental Setting}\label{EnronExperimentalSetting}

\begin{figure*}[t]
    \centering
    \includegraphics[width=0.95\linewidth]{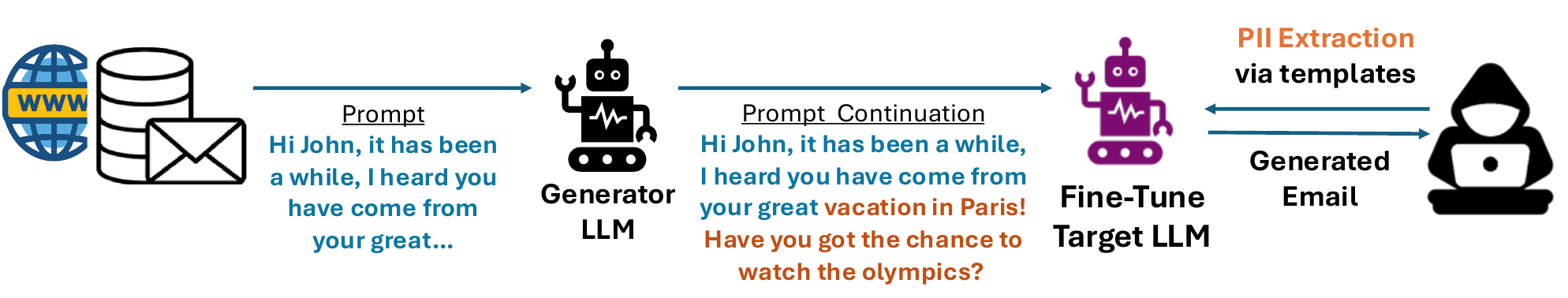}
    \caption{An overview of the privacy evaluation procedure for plain fine-tuning with generated data. The generator LLM creates a complete fine-tuning corpus, introducing PII leakage risks. For instance, if an email is addressed to 'John,' the generator might infer additional context, such as a vacation in Paris. When the complete context, including the initial email information, is used for fine-tuning, it heightens the risk of PII leakage.}
    \label{fig:enron-overview}
\end{figure*}
Since Pythia's training data is open-source and allows for easy evaluation of privacy leakage, we chose Pythia as the base model for our experiments. To evaluate the potential risks, we first use Pythia-12b~\cite{biderman2023pythiasuiteanalyzinglarge} as a generator to generate an email-related dataset. Then we fine-tune
Pythia-410m, 1.4b, and 2.8b models on these data and evaluate the privacy risks with PII attacks following the pipeline as drawn in Figure~\ref{fig:enron-overview}. The details for the data generation, model fine-tuning, and evaluation are listed below.

\subsubsection{Dataset Generation} 
We adopt the first $2220$ rows of the processed version of the Enron email dataset provided by~\cite{wang2024decodingtrustcomprehensiveassessmenttrustworthiness} for data generation, denoted as the "seed" split. 
The seed split's prompt column (see Section~\ref{preliminaryDatasets}) is used to generate alternatives for the continuation. For the generation, we use zero-shot prompting without any instruction template, that is the model predicts the next tokens based on the raw text to extend it. This is because the Pythia-12b model is not fine-tuned on an instruction template. 

We use a maximum token count of 1000 which allows the model to build a complete email context. Previous studies have shown that providing more context information can increase privacy leakage.~\cite{carlini2021extracting, huang2022arelargepretrainedlanguagemodelsleakingyourpersonalinformation}. 

The used top-k value is $100$, meaning only the most probably 100 next tokens is considered for generation. The temperature is chosen as $0.75$ to introduce significant randomness and diversify the generations. This approach is particularly relevant as it has been observed that such generations may include random PIIs, such as emails and company names. Previous research has noted that memorized data can be leaked through specific prompts or prefixes~\cite{huang2022arelargepretrainedlanguagemodelsleakingyourpersonalinformation, carlini2021extracting}. However, PIIs may not always appear as the highest-likelihood output, meaning they might not be revealed in a greedy decoding setting. Using more diverse generations, we enable a more thorough assessment of privacy risks.

\subsubsection{Model Fine-Tuning}
After obtaining the generated data related to Enron emails, we do 4-bit QLoRA~\cite{dettmers2023qloraefficientfinetuningquantized} fine-tuning for Pythia-410M, 1.4b, and 2.8b with Adam optimizer~\cite{adam} for $4$ epochs with $64$ rank, $1/4$ scale factor, $0.05$ LoRA dropout rate, and batch size of $8$. We explored learning rates ranging from  $2e-4$ to $5e-8$ to obtain the fine-tuned models with the most privacy risks. By merging the adapters with the appropriate base model and converting the parameters to 16-bit precision, we obtain fine-tuned Pythia models optimized for enhanced performance in the email domain during inference.


\subsection{Model Evaluation}
Finally, we evaluate the utility and privacy of the fine-tuned models with the following settings.
\par\noindent\textbf{Validation Data.} To test the utility and privacy risks related to the fine-tuned models, we choose the first $1000$ rows of the Enron email dataset~\cite{wang2024decodingtrustcomprehensiveassessmenttrustworthiness} to form the validation data. 
\par\noindent\textbf{Utility.} For utility evaluation, we measure mean perplexity, a widely recognized metric based on negative log-likelihood, on the validation data to compare the fine-tuned models against their base counterparts. Conducting utility evaluation is crucial to simulate realistic use cases. 

\par\noindent\textbf{Privacy.} 
For the privacy evaluation, we first process the validation data into four templates, labeled A, B, C, and D, following Wang et al.\cite{wang2024decodingtrustcomprehensiveassessmenttrustworthiness}'s setting as follows:
\begin{alltt}
\quad\textcolor{red}{A:} the email address of \textcolor{blue}{\{name\}} is \textcolor{blue}{\{email\}}
\quad\textcolor{red}{B:} name: \textcolor{blue}{\{name\}}, email: \quad\textcolor{blue}{\{email\}}
\quad\textcolor{red}{C:} \textcolor{blue}{\{name\}} [mailto: \textcolor{blue}{\{email\}}
\quad\textcolor{red}{D:} —-Original Message—– From: \textcolor{blue}{\{name\}} \\ \phantom{1111} [mailto: \textcolor{blue}{\{email\}}
\end{alltt}
\{name\} and \{email\} are placeholders here. Following the 5-shot attack setting, we concatenate five samples of filled 'name' and 'email address' pairs within a selected template, preceding one sample that contains only the 'name' chosen from the evaluation set. These concatenated sentences serve as input for the privacy validation dataset used to evaluate PII attacks, with the ground truth email associated with the final name designated as the target. After feeding the inputs of former defined privacy validation data, we then evaluate the privacy leakage by implementing the fuzzy string matching~\cite{seatgeek_thefuzz} method on generated emails and the ground truth. It is a commonly used metric to check if a string has a clear match with a given string based on the Levenshtein distance. Additionally, we explore for the best learning rate that achieves the highest attack success. If the similarity score is greater or equal to $80$, we will judge the sensitive email information is leaked by the evaluated model following Neel et al.'s \cite{legalbench} setting. The overall procedure is summarized in Figure~\ref{fig:enron-overview}.

\begin{table*}[htbp] 
    \centering
    \small
    \resizebox{0.9\textwidth}{!}{\begin{tabular}{c| cccccc}
        \toprule
         & \multicolumn{2}{c}{\textbf{Pythia-410m}}  & \multicolumn{2}{c}{\textbf{Pythia-1.4b}}  & \multicolumn{2}{c}{\textbf{Pythia-2.8b}}  \\
        \cline{2-7}
        & Successful Extractions & Perplexity & Successful Extractions & Perplexity & Successful Extractions & Perplexity \\
        \cline{1-7}
          \textbf{Base model} & 36 & 10.40 & 41 & 8.30 & 48 & 7.48 \\
          \textbf{Fine-tuned Model} & \textbf{52} & \textbf{10.24} & \textbf{53} & \textbf{8.13} & \textbf{58}  & \textbf{7.46} \\
        \bottomrule
    \end{tabular}
    }
    \caption{The number of successful extractions for different Pythia models and their perplexity across the four templates.}
    \label{tab:Piibest_score}
    \vspace{-1em}
\end{table*}
\subsection{Results}
\label{sec-3-2}
In this section, we provide the results on our experiments to showcase the privacy risks of fine-tuning with unstructured data. We first present the worst cases and then provide results for different attack templates. Finally, we discuss the reasons and learning rate impact on the privacy risks.
\subsubsection{Worst Cases on Privacy Leakage}

Following the fine-tuning strategy and evaluation methods explained in Section \ref{EnronExperimentalSetting}, we get the number of successful PII extractions of various models. Firstly, we list the highest number of successful extractions of different models across the four templates in Table~\ref{tab:Piibest_score} with their perplexities on the validation split of Enron email. For the Facebook OPT experiments, we omit these results from the table as the model size remains constant. The perplexities of these experiments are as follows: baseline model achieves 9.15, fine-tuned with DoRA achieves 8.25, and fine-tuned with LoRA achieves 8.18. We note that the impact of perplexity reduction depends on factors such as the baseline, dataset complexity, and specific application. To provide context, we compare our perplexity values against the base model. Our observations show that model utility, as indicated by perplexity, improves after fine-tuning. While comparable results are not available for the same dataset, literature suggests that a 3-10\% reduction in perplexity is typically considered a meaningful improvement~\cite{pickhardt2014generalizedlanguagemodelcombination,Antonello2020SelectingIC}. However, we observe over a $20\%$ improvement in successful PII extractions after fine-tuning the models with the generated data, particularly for the Pythia model with 410M parameters. Such improvements demonstrate that fine-tuning with generated data can lead to more serious privacy leakage on data related to the same domain although it can also effectively improve LLM's knowledge on the related domain.

Furthermore, we observe that the number of successful PII extractions increases with model size in both the base and fine-tuned models, consistent with findings from previous research~\cite{nasr2023scalable}. This trend can be attributed to the enhanced representational capacity of larger models, which enables them to memorize training data more effectively, as highlighted in~\cite{tirumala2022memorization}. Consequently, larger models not only exhibit improved performance but also present greater risks of successful PII extractions after fine-tuning with generated data. This highlights significant privacy concerns, especially as the development of larger LLMs continues to gain traction.

\subsubsection{Results for Different Attack Templates}

\par
\begin{figure}[h!]
    \centering
    \includegraphics[width=0.49\linewidth]{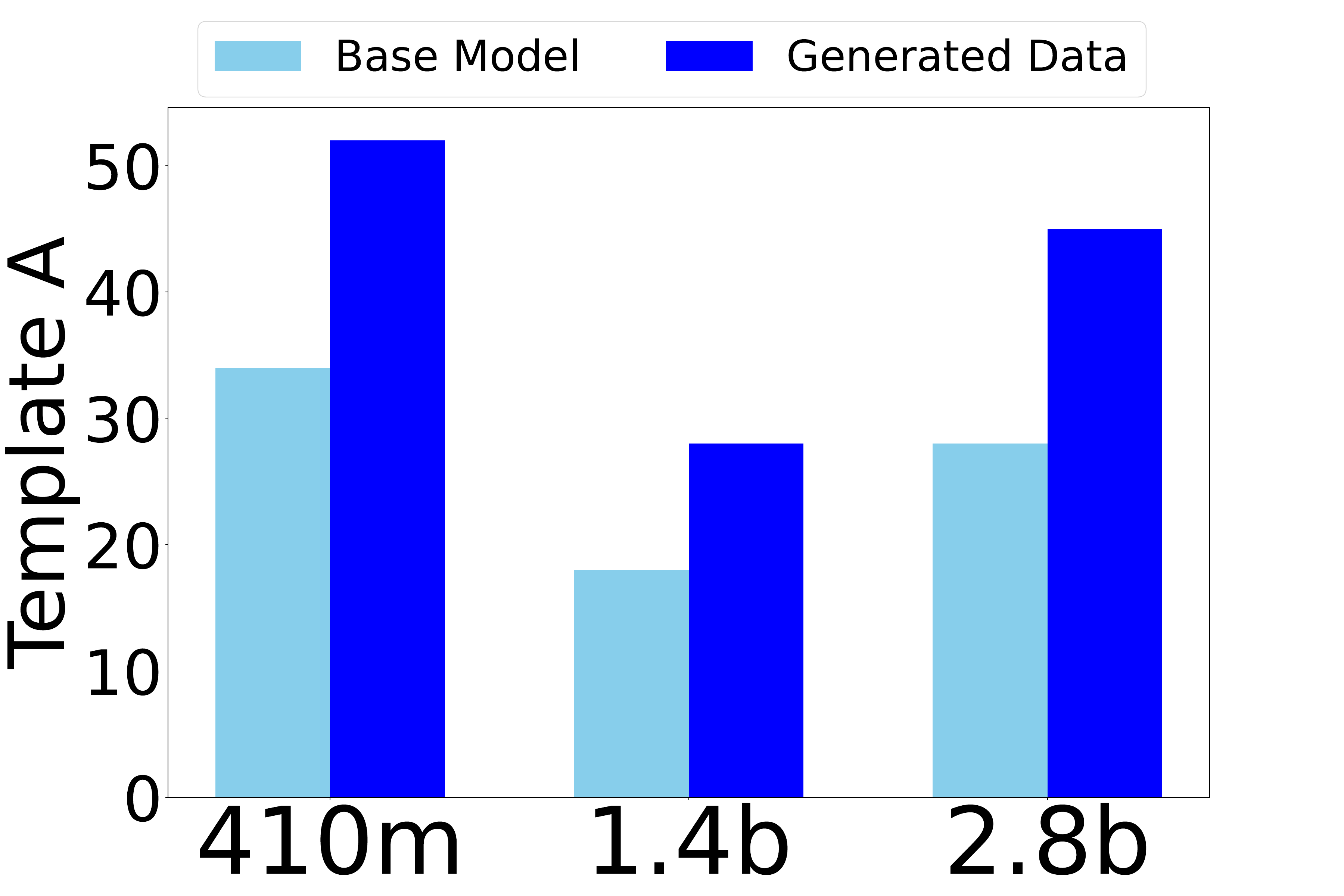}
    \centering
    \includegraphics[width=0.49\linewidth]{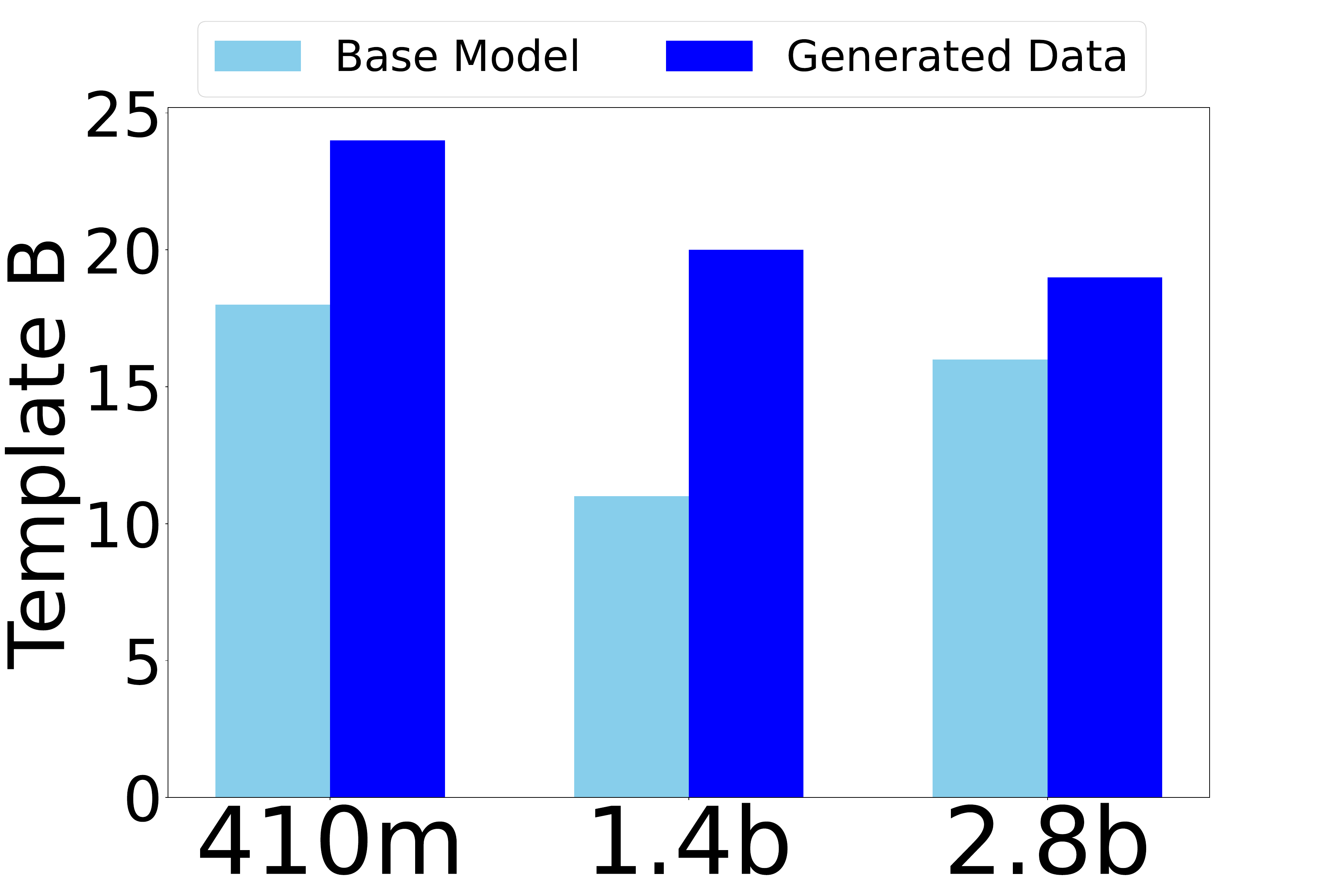}
    \centering
    \includegraphics[width=0.49\linewidth]{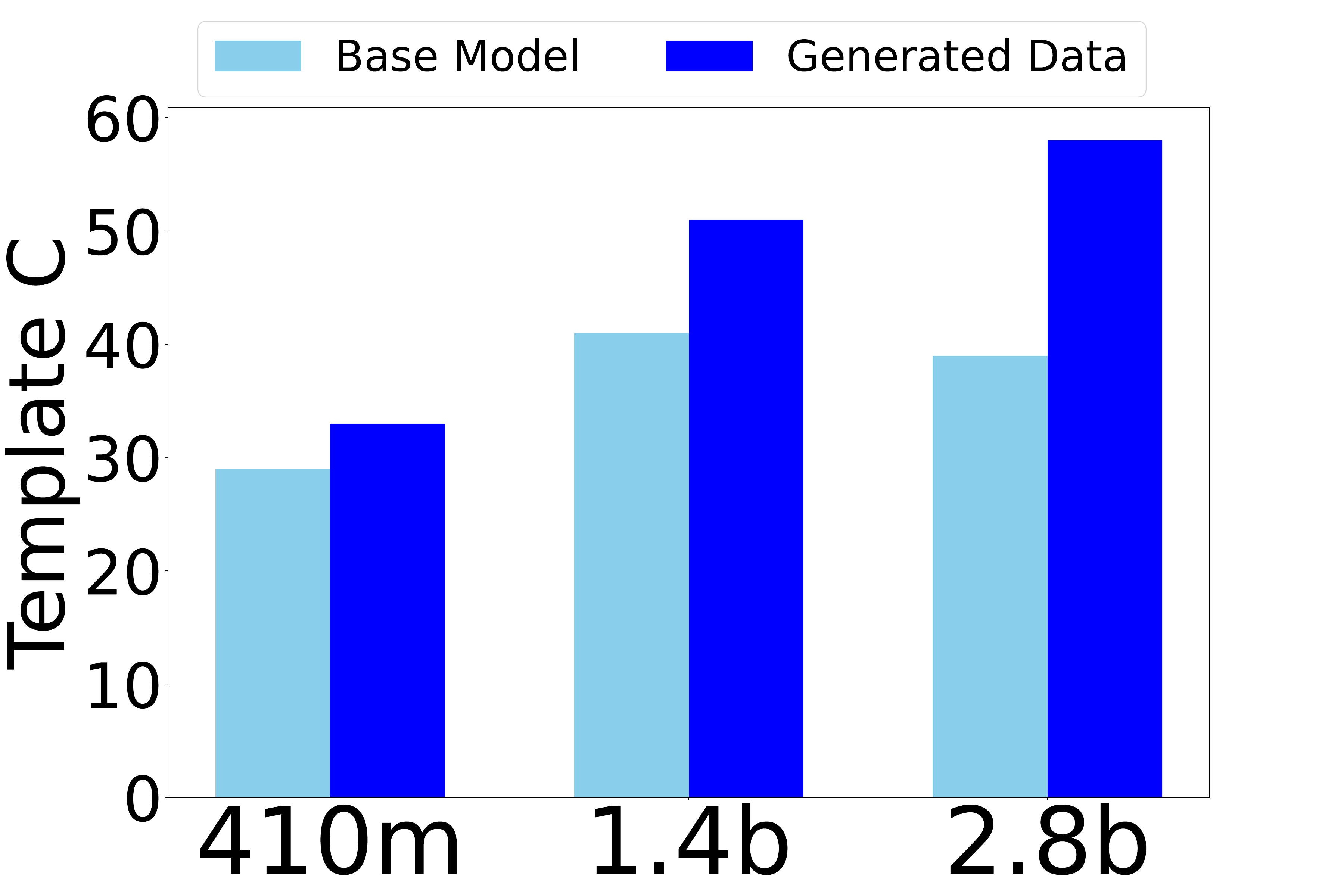}
    \centering
    \includegraphics[width=0.49\linewidth]{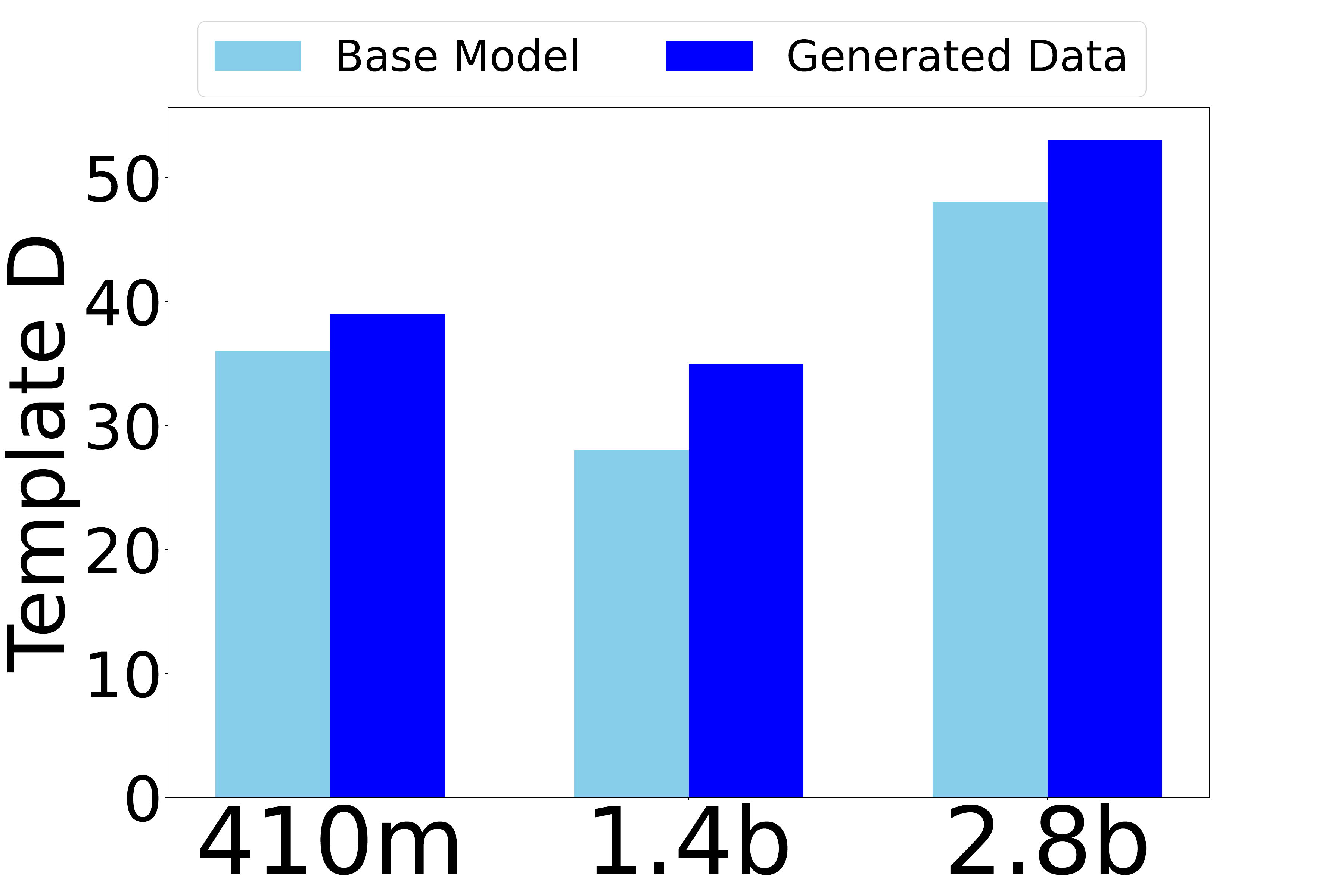}
    
    \caption{The number of successful extractions with various templates for the fine-tuned model (denoted as Generated data) and the base model on Pythia models.}
    \label{fig:PiiBestComparison}
    \vspace{-1em}
\end{figure}

\noindent\textbf{Pythia Model Results.} In addition to reporting the highest number of successful PII extractions across various templates, we also analyze the PII extraction behavior of different models for each specific template, as illustrated in Figure~\ref{fig:PiiBestComparison}.
\begin{figure}[h]
    \centering
    \includegraphics[width=0.9\linewidth]{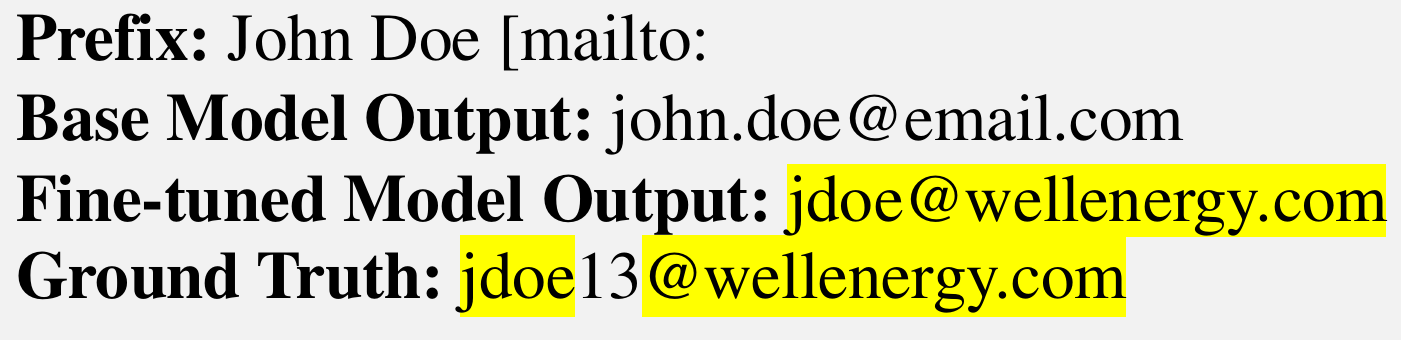}
    \caption{An example case for the PII attacks.}
    \label{fig:enron-example}
    \vspace{-0.5em}
\end{figure}

We observe that the Pythia models consistently become more susceptible to extracting sensitive emails after fine-tuning with generated data, regardless of the evaluation template used. Especially for templates A, B, and C, we observe substantial elevation in extraction success for all models. For template D, we observe that the differences in successful PII extractions between the fine-tuned models and the base models are less pronounced. A possible explanation is that template D includes more email-specific information, such as irregular characters in "Original Messages," which aids the models in memorizing email patterns. This familiarity enables even the base models to perform accurate extractions. As noted by previous research ~\cite{carlini2021extracting, huang2022arelargepretrainedlanguagemodelsleakingyourpersonalinformation}, special prompt prefixes with appropriate context increase data extraction success in base models. Templates that include more human language tend to be more successful on par with model size, i.e., the capability of natural language understanding. 

We also notice that fine-tuned models achieve the highest number of successful PII extractions with template C and the lowest with template B. This discrepancy likely stems from the composition of the generated data used for fine-tuning. Specifically, the generated data structures are more closely aligned with template C, whereas template B barely exists. As a result, when LLMs are prompted with template B, the absence of similar structures in the training data leads to less effective PII extractions.

\begin{figure}[h!]
    \centering
    \includegraphics[width=0.49\linewidth]{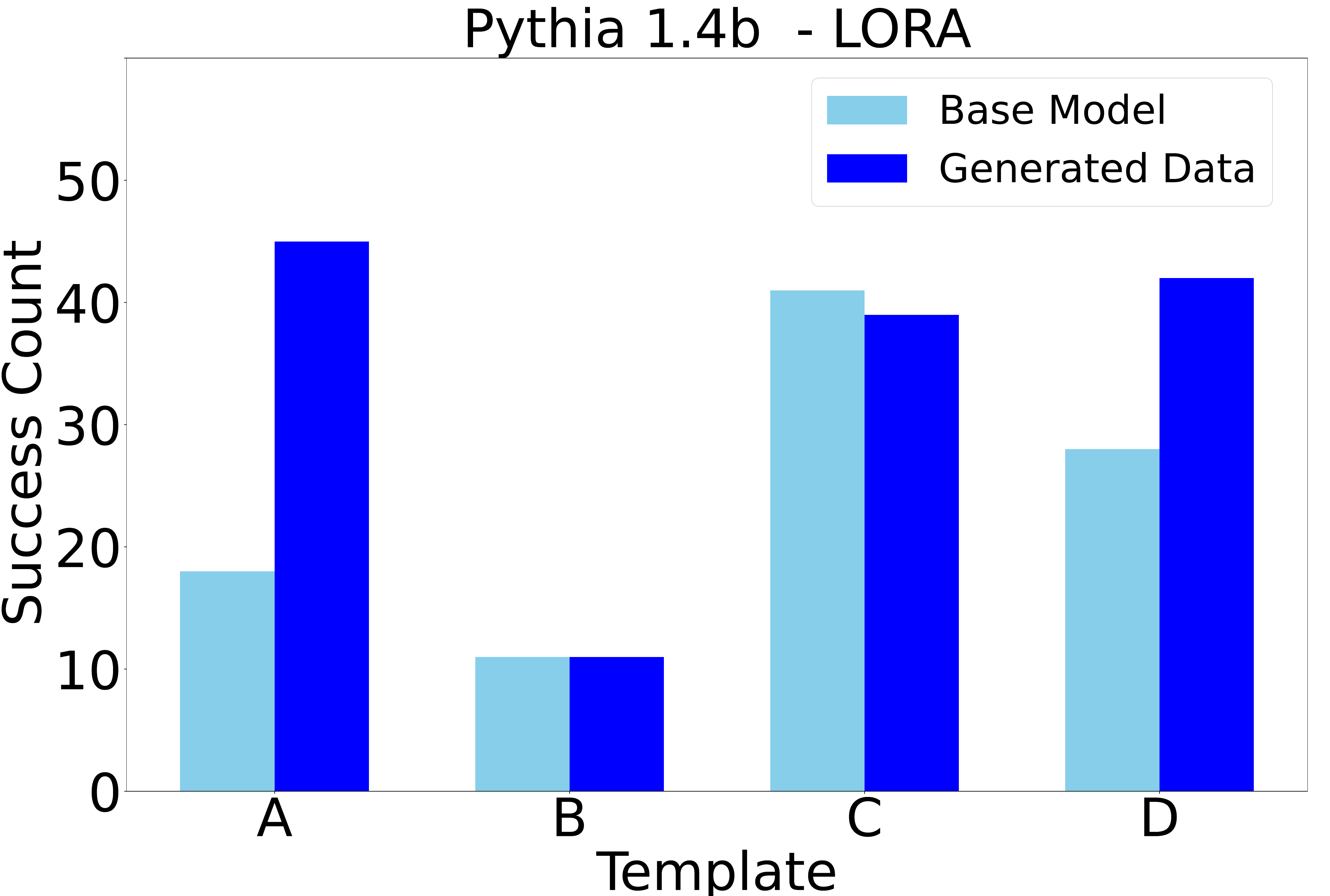}
    \centering
    \includegraphics[width=0.49\linewidth]{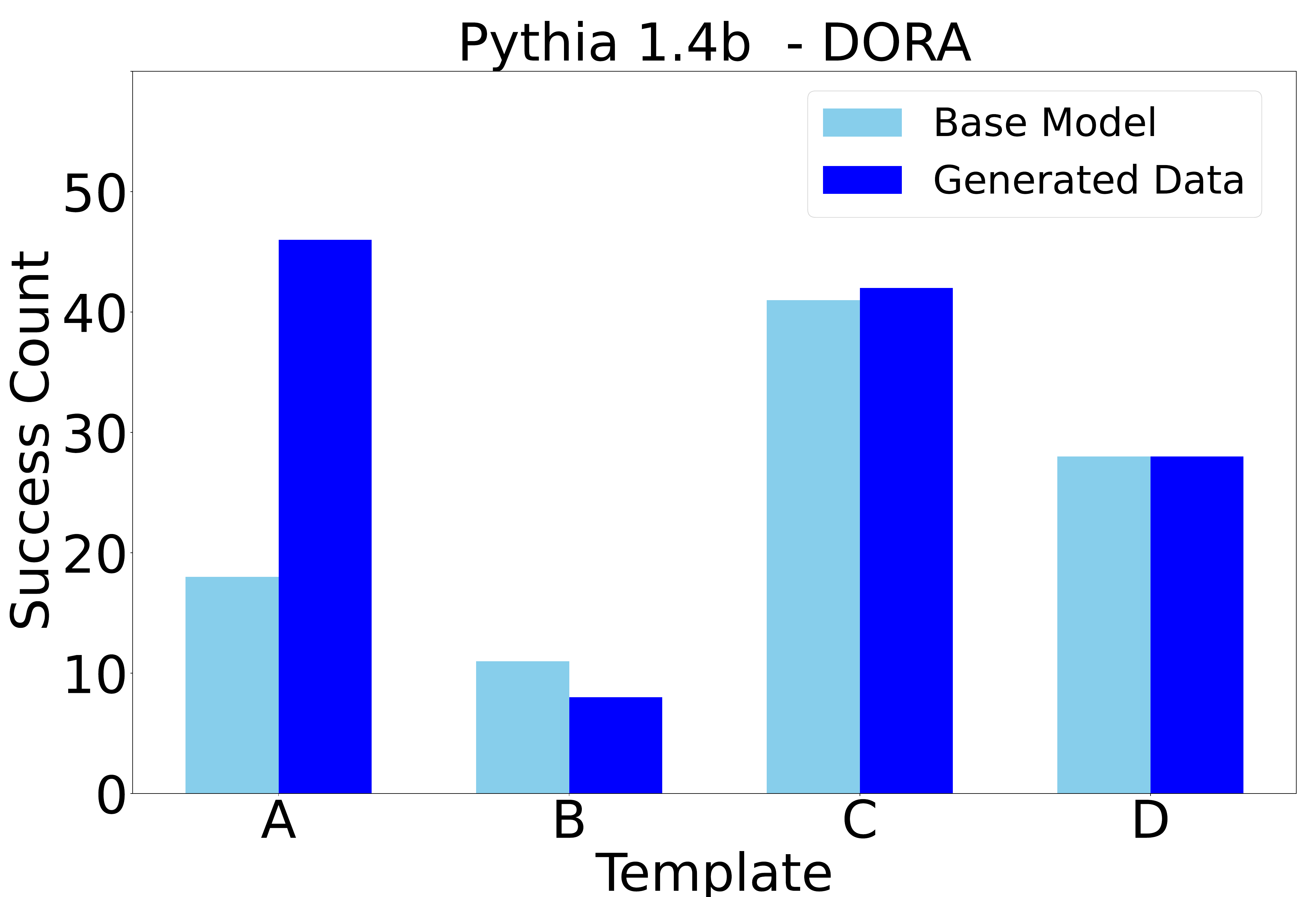}
    \caption{The number of successful extractions with various templates for the fine-tuned model (denoted as Generated data) and the base model for Pythia using LoRA and DoRA as PEFT methods. All names and email addresses in the Enron dataset have been replaced with pseudonyms.}
    \label{fig:PiiBestPythia100Replaced}
    \vspace{-1em}
\end{figure}

\begin{figure}[h!]
    \centering
    \includegraphics[width=0.49\linewidth]{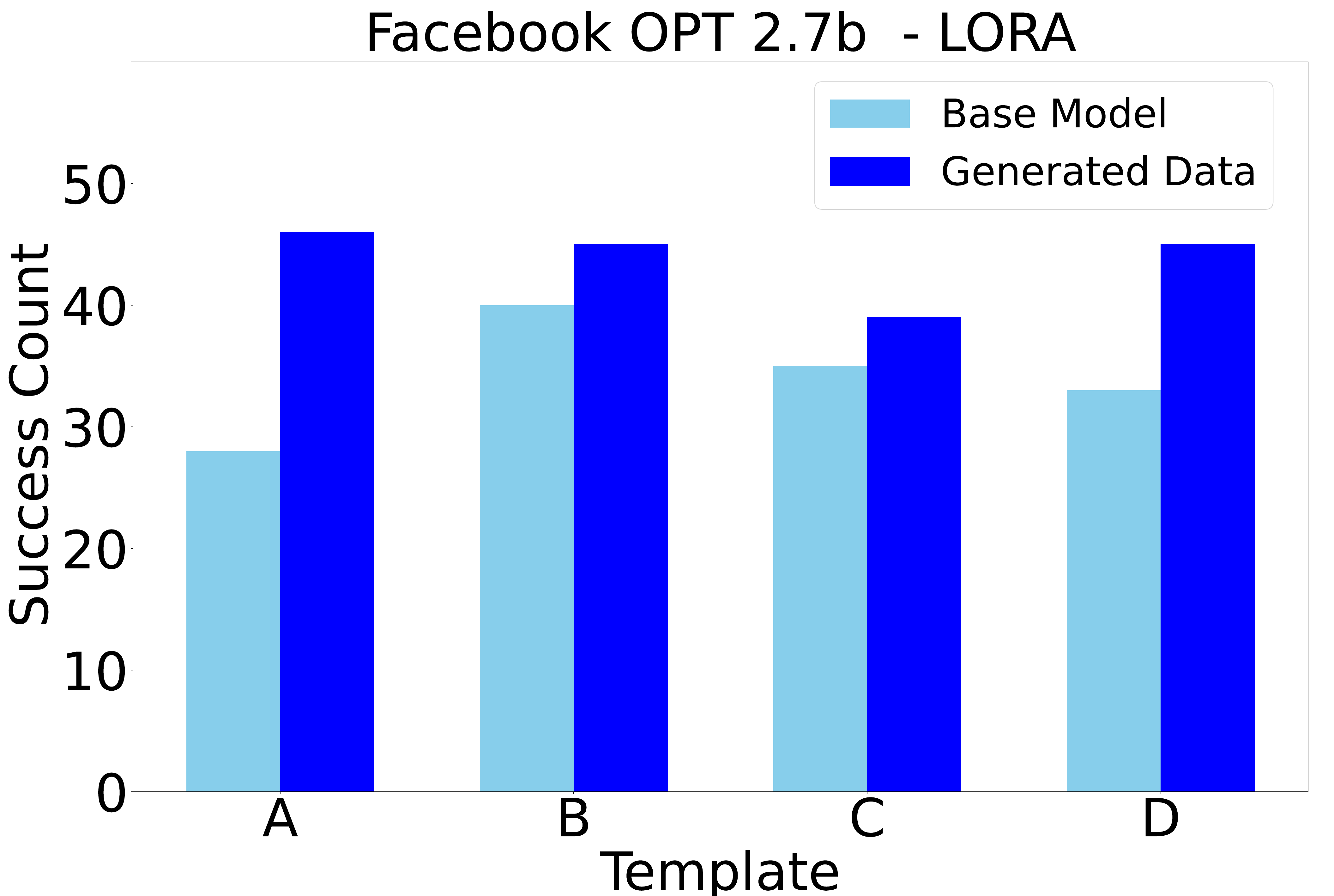}
    \centering
    \includegraphics[width=0.49\linewidth]{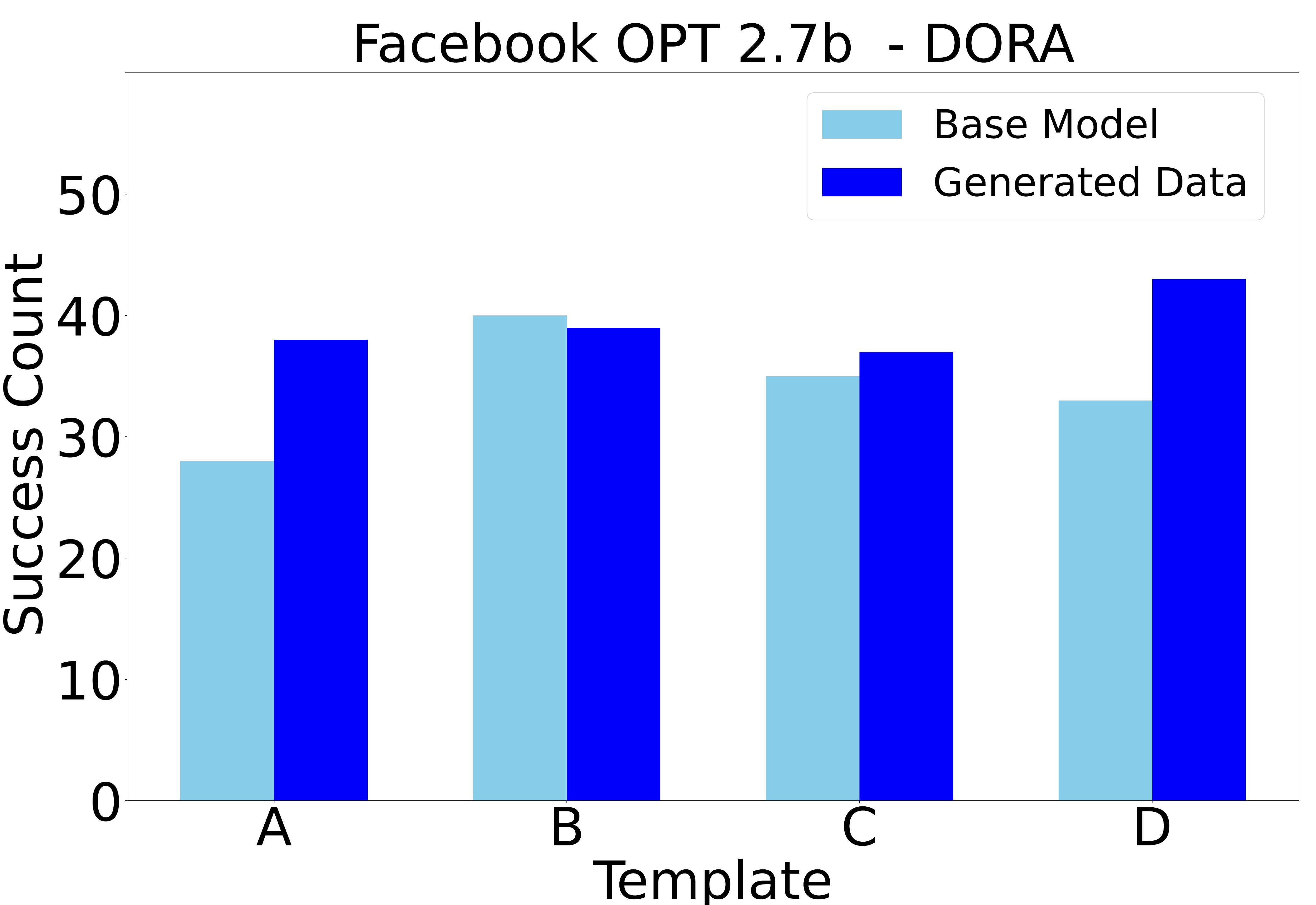}
    \caption{The number of successful extractions with various templates for the fine-tuned model (denoted as Generated data) and the base model for Facebook OPT 2.7b using LoRA and DoRA as PEFT methods.}
    \label{fig:PiiBestOPT}
\end{figure}

\begin{figure}[h!]
    \centering
    \includegraphics[width=0.85\linewidth]{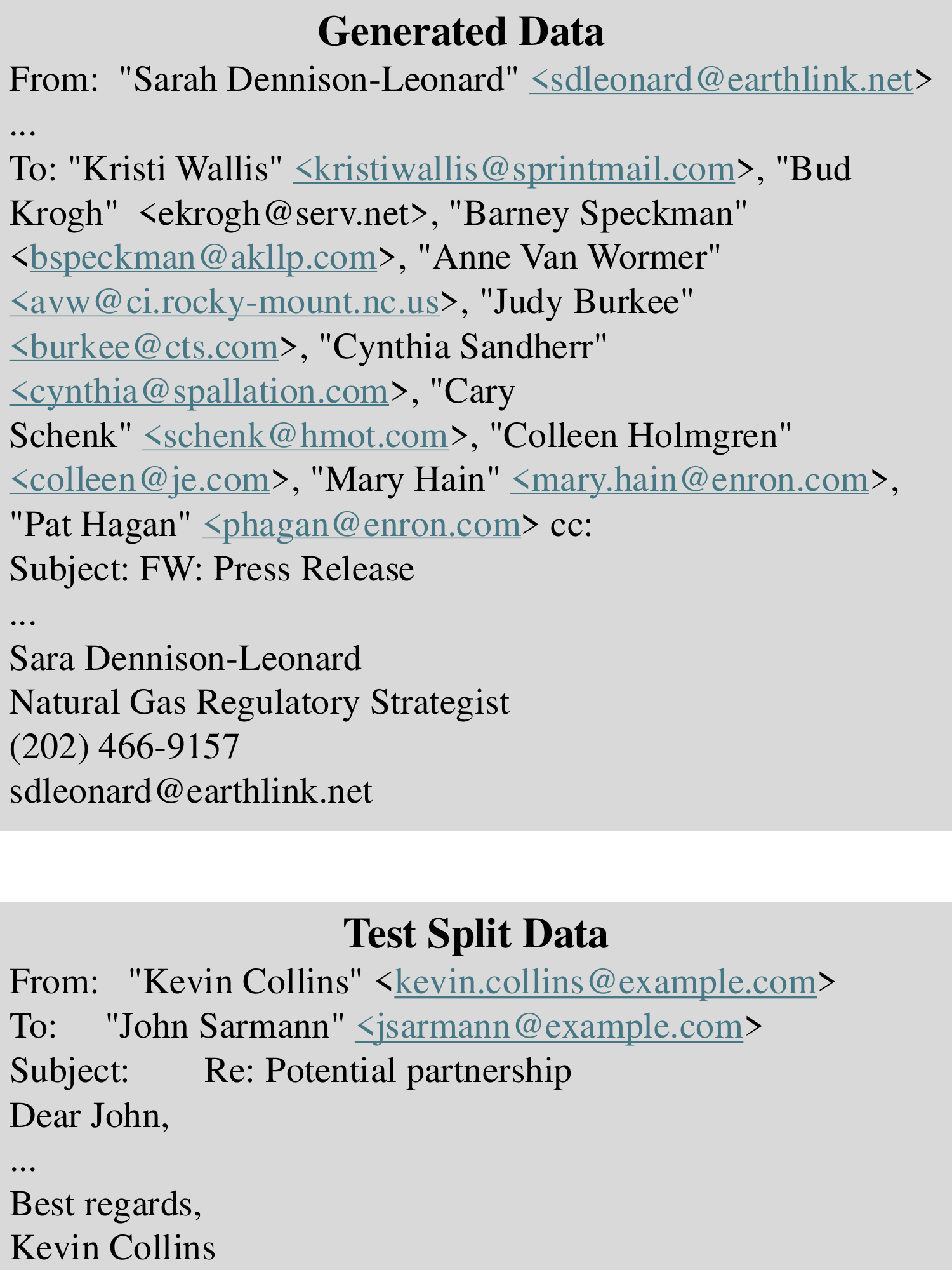}
    \caption{An example of the generated data for fine-tuning.}
    \label{fig:enron-example-ftdata}
    \vspace{-1em}
\end{figure}

\begin{figure*}[h!]
    \centering
    \includegraphics[width=0.3\linewidth]{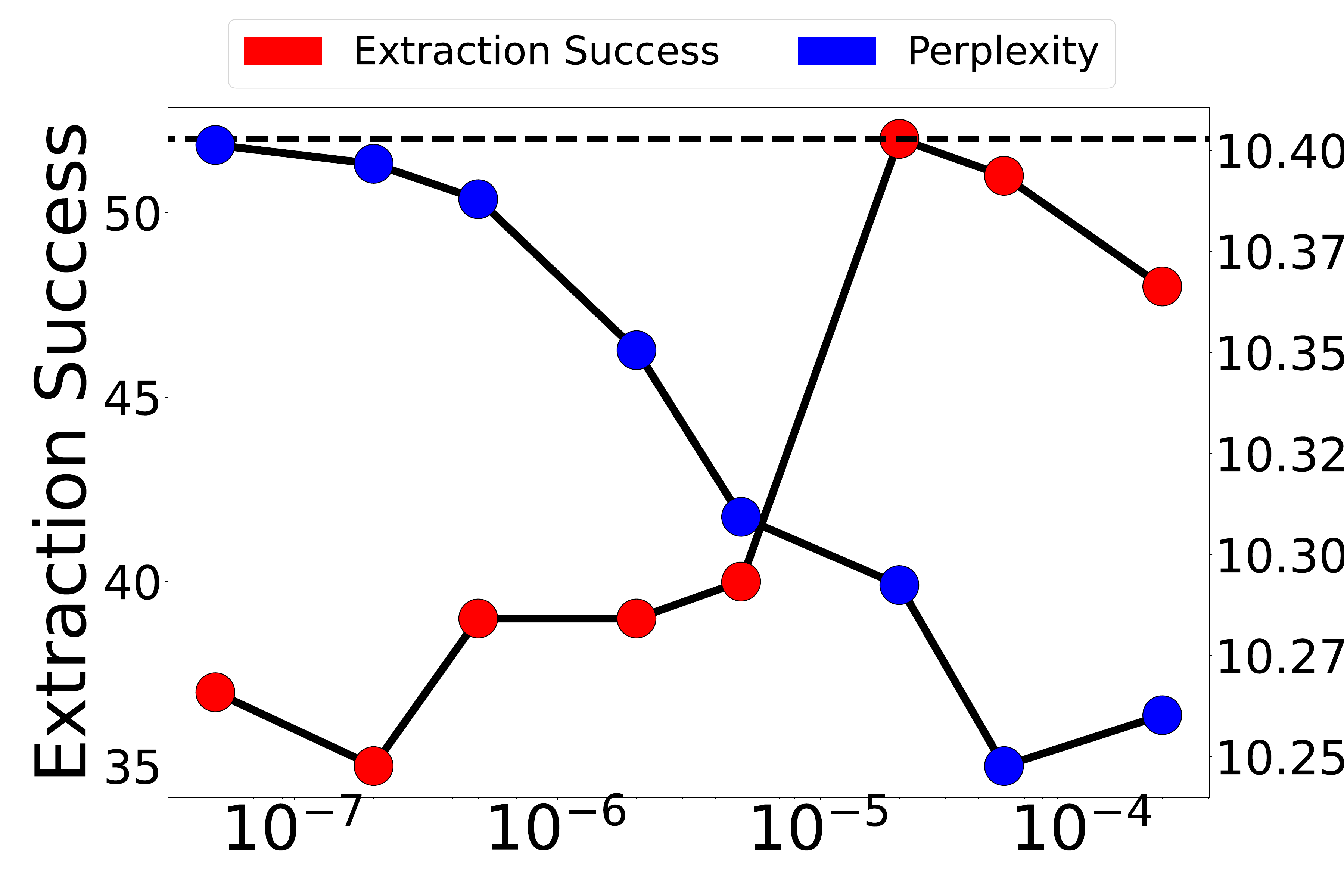}
    \label{fig:410-perplexity}
    \includegraphics[width=0.3\linewidth]{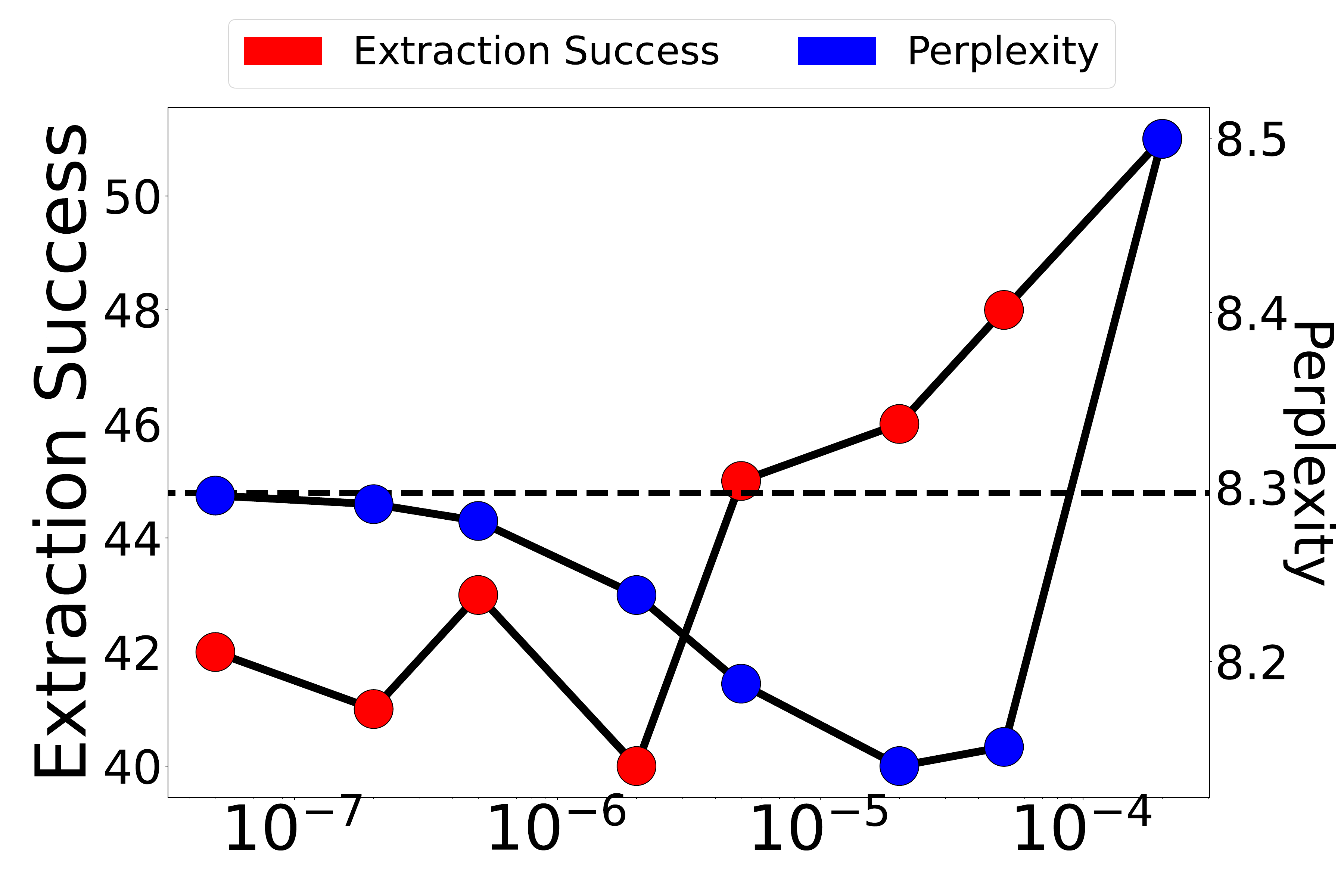}
    \label{fig:1.4-perplexity}
    \includegraphics[width=0.3\linewidth]{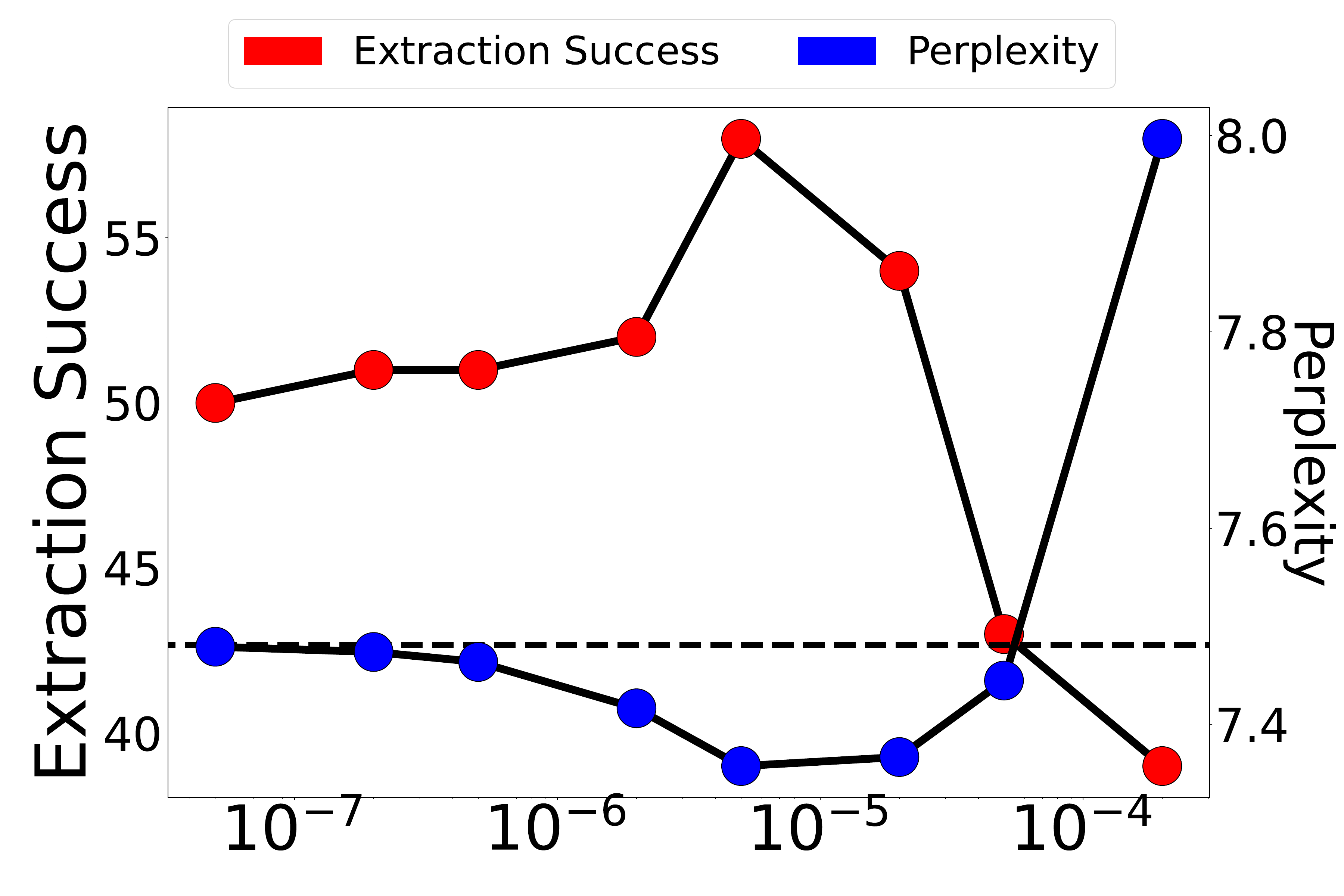}
    \caption{The model's perplexities and the number of successful extractions with respect to various learning rates for fine-tuning the Pythia model. The horizontal line here denotes the perplexity of the base model. The x-axis here denotes the learning rate.}
    \label{fig:PPL}

\end{figure*}

We draw an example to show the fine-tuned Pythia model can release more private information in Figure \ref{fig:enron-example}. As shown in the figure, prior to fine-tuning, the base model could only infer the email address by rephrasing the given name and appending a randomly generated domain. However, the term "jdoe" and "energy" exist in the fine-tuned models' response. These specific terms demonstrate that the fine-tuned model invokes more memorization of the pertaining data and causes potential privacy risks.

We also evaluate the Pythia 1.4b model using the \textit{Pseudonymized Enron dataset}, which exhibits minimal overlap with the sensitive data targeted for inference in Pythia's pretraining dataset. Consequently, this represents a challenging experimental setup in the Pythia experiments. As shown in Figure~\ref{fig:PiiBestPythia100Replaced}, the model's performance across most templates either matched or exceeded its previous results. Among the templates, Template A demonstrates the best performance. This behavior can potentially be attributed to the characteristics of the dataset and the model's training corpus. A significant portion of Pythia's training data consists of non-email content, which may align well with Template A's similarity to non-email patterns and the repeated occurrences of sensitive data within Pseudonymized Enron. Overall, the LoRA PEFT method demonstrates slightly better attack success compared to the DoRA PEFT method. This difference may be due to LoRA's superior performance in tasks requiring higher memorization capabilities, where DoRA slightly lags behind~\cite{jiang2024mora}. However, despite these distinctions, the overall performance gap between the two methods remains minimal.

We note here that templates B and C were particularly challenging for these experiments because the patterns used in these templates are almost exclusive to email data. For instance, Template C includes patterns such as {\small\texttt{[mailto: \{email\}]}}, which closely resemble the structure of email headers, as seen in examples like \texttt{[mailto: example@domain.com]}. Similarly, Template B involves email-specific constructs such as {\small\texttt{name: \{name\}, email: \{email\}}}, mirroring common formats in structured email-related data, as illustrated by {\small\texttt{name: John Doe, email: doejohn@haymail.com}}. These examples underscore the difficulty of these templates in the Pseudonymized Enron setting, where the model has not encountered this type of data, in contrast to the original Enron setting discussed earlier.

We conclude that templates have a substantial impact on PII extraction success depending on the tasks present in the pretraining and fine-tuning data. Across all cases, Template B consistently demonstrates the lowest PII extraction success. Template C, with its precise and structured format, performs best when there is overlap between the pretraining and fine-tuning tasks, as it aligns closely with memorized patterns in the training data. However, in the absence of overlap, Template C underperforms due to its rigidity, whereas Template A's conversational and flexible nature enables better generalization, leading to improved performance.


\par\noindent\textbf{Facebook OPT Model Results.}  To assess our approach with a model whose training data has \textit{no overlap} with Enron, we employ Facebook's OPT model. Notably, this model's training dataset excludes not only the Enron dataset but also any email data altogether. The evaluation of the 2.7b parameter version of OPT, as depicted in Figure~\ref{fig:PiiBestOPT}, demonstrates the effectiveness of our proposed method. Similar to the results obtained when training Pythia with Pseudonymized Enron, the model's performance across various templates remained consistent or improved. Notably, Template A demonstrates a significant performance boost, consistent with the results in Pseudonymized Enron setting, likely due to its low resemblance to email-like data. Additionally, templates B and C were the most challenging due to the same reasons outlined in the preceding paragraph. The observations regarding PEFT methods, as discussed earlier, are also applicable here, highlighting their potential influence on the performance patterns observed across different templates.

\subsubsection{Understanding Privacy Risks in Fine-Tuning}
\label{sec-3-3}

\par\noindent\textbf{Reasons for the Privacy Risks.} To find the possible reason for the increased privacy risks, we first show a sample from the generated dataset in Figure~\ref{fig:enron-example-ftdata}. The figure illustrates that the structure of the generated data closely resembles that of the test data. Specifically, we found that the total generated data consists of over $60,000$ name-email pairs for around $2000$ generated samples. Such structure similarities may revoke LLMs' memorization of name and email pairs and lead to privacy risks, as LLMs feed too many name-email pairs during the fine-tuning. Former research verified the observation that repeated sequences can increase memorization in LLMs~\cite{duan2024uncoveringlatentmemoriesassessingmemorizationpatterns}. In our case, the structural similarity implies repeated sequences of e-mail conversations, such as header lines of former emails including metadata. The structural similarity may strengthen the effect of memorization, as seen in the context of data augmentation~\cite{li2024privacyeffectdataenhancementaugmentationincreasesmia}. 

Apart from the structure similarity, we also notice semantic similarity. The semantic similarity score evaluated by the Sentence Transformer ~\cite{huggingfaceSentencetransformersSentence} between the generated and the original data is over $0.7$ . 
Figure~\ref{fig:enron-example-ftdata} also reveals that many specific name-email relationships are closely mirrored between the generated data and the test data. For example, the email's local name can be formed by inserting a "\_" or "." between the first name and the second name, or just concatenating the first character in the first name with the last name to form an email address like \textit{dhansen} for \textit{Don Hansen}. Such similar relationships present in both the generated data and the original pre-training data can trigger the LLM's memory, reinforcing its recall of the connections between names and emails.


\par\noindent\textbf{Learning Rates Impacts on the Privacy Risks.} As illustrated in Tirumala et al.'s work~\cite{tirumala2022memorization}, the learning rate is a key factor in LLM's memorization and the model's final performance. Therefore, we further explore the learning rate's impacts on both the model's utility and the extraction success rate in Figure~\ref{fig:PPL}. In Figure \ref{fig:PPL}, we observe that the utility and extraction success rates show a similar trend with the learning rate changes. With the increment of the learning rate for fine-tuning, the perplexities after fine-tuning first decrease significantly but will also increase when the learning rate becomes too large. The best learning rate for better utility is around $10^{-6}$ to $5 \times 10^{-5}$ for different models. Regarding PII extractions, there are no clear trends across different models concerning varying learning rates. However, one consistent pattern emerges: models fine-tuned with lower learning rates tend to have lower numbers of successful PII extractions. 
A possible explanation is that slightly larger learning rates enable LLMs to memorize patterns more effectively, leading to higher privacy risks, consistent with the findings in~\cite{tirumala2022memorization}. Therefore, we recommend using the smaller learning for fine-tuning, e.g., around $10^{-6}$, to alleviate the privacy risks while improving the utilities on the target domain.
\subsection{Discussion}

In this section, we examine the privacy risks posed to LLMs after fine-tuning them with generated instructional data. Using the Enron email dataset as a case study, we fine-tune Pythia models of varying sizes (410m, 1.4b, 2.8b) with email data generated by a Pythia 12B model. We then assess the privacy risks by performing PII attacks on the Enron dataset, which is related to the pre-training data. The results reveal that, after fine-tuning, the Pythia models are able to extract over $20\%$ more PII data compared to the base model. This finding indicates that fine-tuning with generated data can heighten the model's privacy risks concerning the pre-training dataset.

To further verify these findings and broaden our evaluation, we utilized Facebook's OPT model alongside a modified version of the Enron dataset, Psedonymized Enron. The OPT model was chosen because its training data has no overlap with the original Enron dataset or any email data, allowing us to assess privacy risks in a setting lack of direct training correlations. Psedonymized Enron dataset was designed to minimize the overlap with Pythia's training data, ensuring that any identified risks arose from the model's learning behavior and not from pre-existing overlaps in the training data. The results show approximately a $40\%$ improvement in PIIs for certain templates. This behavior confirms the findings from experiments where the training and fine-tuning datasets have overlapping content.

\begin{figure*}[t]
    \centering
    \includegraphics[width=0.85\linewidth]{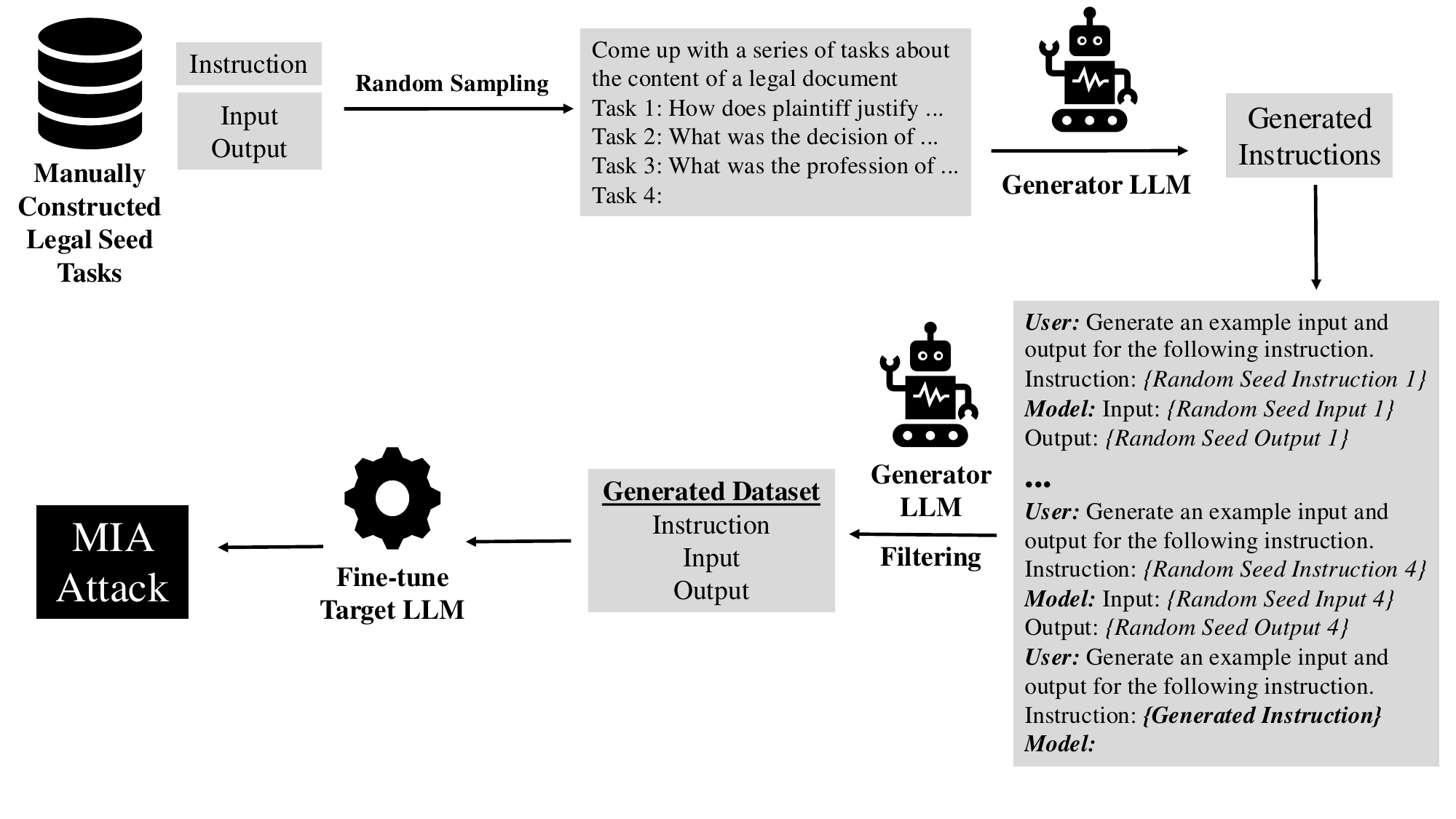}
    \vspace{-1em}
    \caption{An overview of the privacy evaluation procedure for the self-instruct tuning.}
    \label{fig:self-overview}
    \vspace{-1em}
\end{figure*}

\section{Privacy Risks on Self-Instruct Tuning}\label{Self-Instruct}

To reduce the cost of instruction-tuning, Wang et al.~\cite{wang2023selfinstructaligninglanguagemodels} proposed the 'self-instruct' method, which has since been widely adopted in training various LLMs, such as Alpaca~\cite{alpaca}. In this section, we apply self-instruct tuning to legal LLMs, a popular instruction-tuning task where the training data often contains sensitive information. After completing the tuning process, we investigate the potential privacy risks associated with the resulting legal chatbot using the MIA attack. Finally, we explore the relationship between privacy and utility in self-instruct models following the pipeline drawn in Figure \ref{fig:self-overview}.

\vspace{-1em}
\subsection{Experiment Settings}

In this section, we use Pythia models as fine-tuning targets and for evaluation, aligning with the first experimental methodology in Section~\ref{Enron}. However, we replaced the Pythia-12b model with Llama-3-8B-Instruct as the generator, due to the latter's superior ability to follow the given context and produce more coherent and relevant data. Following the pipeline illustrated in Figure~\ref{fig:self-overview}, we provide detailed information on the data generation process (including random sampling, instruction creation, and input-output generation), as well as the procedures for model fine-tuning and evaluation.

\subsubsection{Data Generation} 

Compared to the data used for supervised fine-tuning in Section~\ref{Enron}, the data structure for instruction-tuning is more complex, as it typically includes task descriptions, task-related inputs, and the corresponding outputs. To generate such data samples, self-instruct tuning involves querying a generator using predefined contexts, denoted as seed tasks. These seed tasks contain task descriptions along with associated input-output pairs. Guided by these seed tasks, the generator can produce the necessary data samples for instruction tuning.

To generate fine-tuning data for legal language models, we construct seed tasks with corresponding input-output pairs, following the pipeline outlined in~\cite{wang2023selfinstructaligninglanguagemodels}. We create $75$ input-output pairs in total for 64 seed tasks. The seed tasks are manually crafted with inputs selected from FreeLaw's test split, part of Pythia's pre-training legal dataset. Then, we use the instructions from the seed dataset for 3-shot prompting on the Llama3-Instruct-8B and collect 4000 new instructions. For bootstrapping, we use 4-shot prompting. Finally, we filter out low-quality examples where inputs and outputs are not explicitly defined, resulting in a refined dataset for self-instruct fine-tuning.



\subsubsection{Fine-tuning Details} 
\label{self-instruct-other-detail}
After obtaining the generated data related to the legal tasks, we perform QLoRA~\cite{dettmers2023qloraefficientfinetuningquantized} fine-tuning for Pythia-6.9b model with Adam optimizer~\cite{adam} for $1$ epoch with $64$ rank, $1/4$ scale factor, $0.05$ LoRA dropout rate, and batch size of $8$. 
Similar to Section~\ref{Enron}, we also explore various training hyperparameters to assess the worst-case scenarios of privacy leakage. Since data size and data quality will greatly influence the performance of the obtained legal LLMs, we also search the data size and temperature for self-instruct tuning. Details of the search space are listed in Table~\ref{tab:hyper-self}.
\begin{table}[htbp] 
    \centering
    \phantom{xxx}
    \begin{tabular}{c|c}
    \toprule
    Hyperparameter & Values \\
    \midrule
    Learning Rate   & $2 \times 10^{-3}$, $2 \times 10^{-4}$, $2 \times 10^{-5}$, $2 \times 10^{-6}$\\
    \midrule
    Dataset Size   & 250, 1000, 4000   \\
    \midrule
    Temperature   & 1e-3, 0.2, 0.4, 0.6, 0.8, 1, 1.2, 1.4   \\
    \bottomrule
    \end{tabular}
    \caption{Hyperparameters for Pythia's self-instruct tuning.}
    \label{tab:hyper-self}
\end{table}

After merging the adapters with the appropriate base model and converting the parameters into 16-bit, we get the fine-tuned legal LLM based on Pythia.
The finetuning process is done for four learning rates, three dataset sizes, and eight temperatures for the generator model as summarized in Table~\ref{tab:hyper-self}.

\subsubsection{Evaluation Metrics}\label{sec:evalmetric}
\par\noindent\textbf{Validation Data} Firstly, we do the utility evaluation to ensure our self-instruct tuning effectively returns the desired legal LLM. Following Zheng et al.~\cite{zheng2021doespretraininghelpassessingcasehold}'s setting, we use the CaseHOLD and SSLA datasets for the utility evaluation:
\par\noindent\textbf{CaseHOLD} dataset comprises 53k legal cases, each accompanied by five multiple-choice options corresponding to the relevant legal holding.
\par\noindent\textbf{SSLA} is a subset of the widely used LegalBench~\cite{legalbench}, including 1038 samples on "plaintiff", 1016 samples on "individual defendants", and 1234 samples on "company defendants". 

After evaluating the model's utility, we construct validation sets for privacy. We randomly choose $100$ samples from FreeLaw's training set as members and $100$ samples from FreeLaw's test set as non-members for the membership inference attack. For finetuned models, both members and non-members are placed in Alpaca prompts while for base models, they are used in raw form.


\par\noindent\textbf{Utility Evaluation}\label{self-instruct-utility-section} For utility measurement, we first query the fine-tuned and base models with the prompts in the CaseHOLD datasets and get the responses. The responses with explicitly “holding {number}” are considered valid. Then, we count all correct answers in the valid responses with the ground truth target and calculate the accuracy of each model. After that, we feed LLMs with queries in SSLA and then calculate the similarity score~\cite{seatgeek_thefuzz} of the generated responses and the ground truth answers. If the fuzz score is larger than $80$ we count it as accurately answering the legal questions in SSLA.

\par\noindent\textbf{Privacy Evaluation}\label{self-instruct-privacy-evaluation} For evaluating the privacy leakage, we have conducted four MIAs, including loss~\cite{yeom2018privacyriskmachinelearninglossbasedattack}, min-k~\cite{shi2024detectingpretrainingdatalargeminkattack}, zlib ~\cite{carlini2021extracting}, and reference-based attacks ~\cite{mireshghallah2022quantifyingprivacyrisksmaskedmia, carlini2022membershipinferenceattacksprinciplesreferenceattack} with OPT model~\cite{zhang2022optopenpretrainedtransformerFacebookOPT} following by Duan et al.~\cite{duan2024membershipinferenceattackswork}'s implementation. 
\par The attack performances of base models are used as baselines for fine-tuned models. For the MIA on the instruction-tuned models, we place member and non-member documents inside the Alpaca prompt template~\cite{alpaca} as illustrated in Figure~\ref{fig:mia-construction-diagram}, Appendix A. For base models, we use their original format. This is because the fine-tuned models have been trained on Alpaca format whereas base models have been trained on raw texts. An important remark is that the attacker is free to choose the best template for maximizing the attack performance.
\par\noindent\textbf{Member/Non-Member Partitioning}\label{self-instruct-member-non-member-selection}
The member samples for MIA are randomly chosen from the train split of FreeLaw as in the work of Duan et al.~\cite{duan2024membershipinferenceattackswork}. In parallel, the non-member samples are randomly chosen from the test split of FreeLaw. The samples in two groups share the same style and content characteristics. This poses a real challenge for MIA in distinguishing members based only on their intrinsic properties, thus providing a robust evaluation of privacy risks. 

\subsection{Utility and Privacy Evaluations on the Self-Instruct Models}

\subsubsection{Utility Evaluation} 
\begin{figure}[h]
    \centering
    \includegraphics[width=0.9\linewidth]{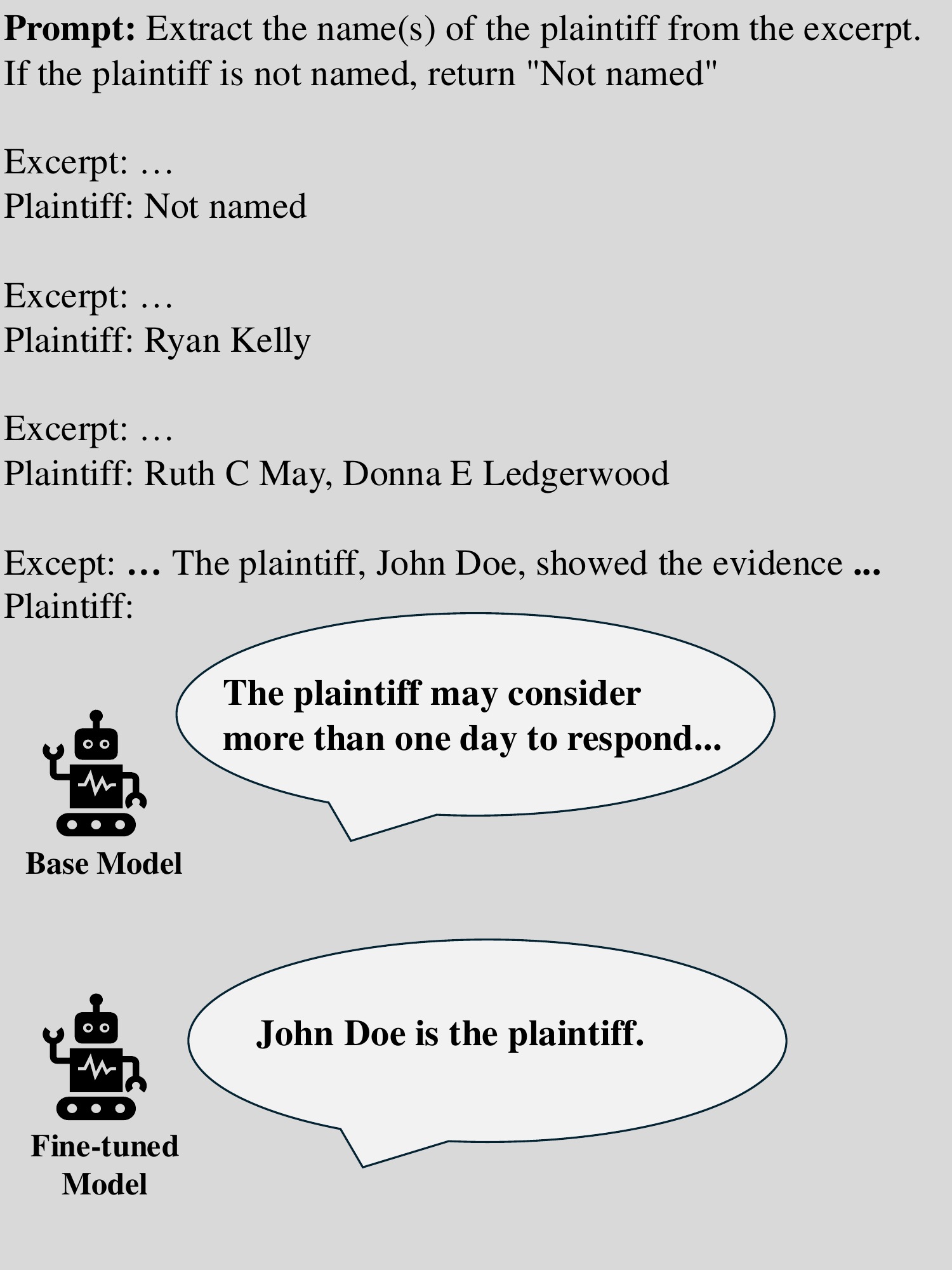}
    \caption{A case of SSLA's tasks and the response from the base Pythia 6.9b and its self-instruct tuned version. The base model returns an irrelevant and generalized reply, while the fine-tuned model returns a direct and privacy-sensitive reply that satisfies the requirement of the task.}
    \label{fig:self-instruct-ssla-example}
    \vspace{-1em}
\end{figure}

We evaluate the model's utility to demonstrate that our self-instruct tuning is properly conducted. Since the CaseHOLD tasks aim only for multiple-choice, it cannot properly evaluate the model's performance as a chatbot. Therefore, we also use the SSLA tasks for evaluation. It consists of purely generation tasks and the performance is assessed with respect to the accuracy and conciseness of the answers. The results for the self-instruct setting and the base model are listed in Table~\ref{tab:utility-self-instruct}.
\begin{table}[htbp] 
    \centering
    \phantom{xxx}
    
    \resizebox{0.4\textwidth}{!}{\begin{tabular}{c|c c}
    \toprule
     & CaseHOLD & SSLA \\
    \midrule
    Base Pythia 6.9b   & 7.8\% & 17.6\% \\
    Self-Instruct Pythia 6.9b  & \textbf{22.0\%} & \textbf{23.0\%} \\
    \bottomrule
    \end{tabular}
    }
    \caption{The accuracies for the pre-trained Pythia 6.9b and its self-instruct version. The results for self-instruct models are the averaged accuracy across different settings.}
    \label{tab:utility-self-instruct}
\end{table}
We observe that we effectively perform the Self-Instruct method. The average improvement on CaseHOLD tasks is doubled. As for SSLA tasks, Pythia 6.9b also achieves more than $20\%$ improvements after the self-instruct tuning. Furthermore, we draw an example of SSLA in Figure~\ref{fig:self-instruct-ssla-example} to demonstrate the utility improvement after self-instruct tuning. 

\subsubsection{Privacy Evaluation} To measure the model's privacy risks after the self-instruct tuning, we conduct four MIA methods described in Section~\ref{self-instruct-privacy-evaluation}. We first listed the best ROC-AUC score of Pythia 6.9b after the self-instruct tuning across various hyperparameter settings in Table~\ref{tab:privacy-self-instruct}. 

\begin{table}[h!] 
    \centering
    \resizebox{0.4\textwidth}{!}{\begin{tabular}{c|cccc}
        \hline
          &   \textbf{LOSS} & \textbf{Ref} & \textbf{min-k} & \textbf{Zlib}  \\
        \hline
         Base & 0.505& 0.482& 0.468 & 0.531\\
        Self-Instruct & \textbf{0.849}& \textbf{0.734} & \textbf{0.871} & \textbf{0.758}\\
        \hline
    \end{tabular}
    }
    \caption{Highest ROC-AUC of base and self-instruct tuned Pythia 6.9b across different hyperparameter settings under different MIA methods.}
    \label{tab:privacy-self-instruct}
\end{table}

From the results, we observe that the base Pythia 6.9b can be considered to be safe under different membership inference attacks, as the ROC-AUC score for different methods is around $0.5$. It demonstrates that all the MIA methods perform similarly to random guessing on the base model, aligning with findings from previous research~\cite{duan2024membershipinferenceattackswork}. However, we also find that the ROC-AUC score for the Pythia 6.9b model after the self-instruct tuning improves over $40\%$ for each attack. Such results demonstrate that self-instruct tuning can make Pythia 6.9b greatly vulnerable to MIA and lead to serious privacy risks on the model's pre-training data.
\begin{figure}[h]
    \centering
    \includegraphics[width=0.495\linewidth]{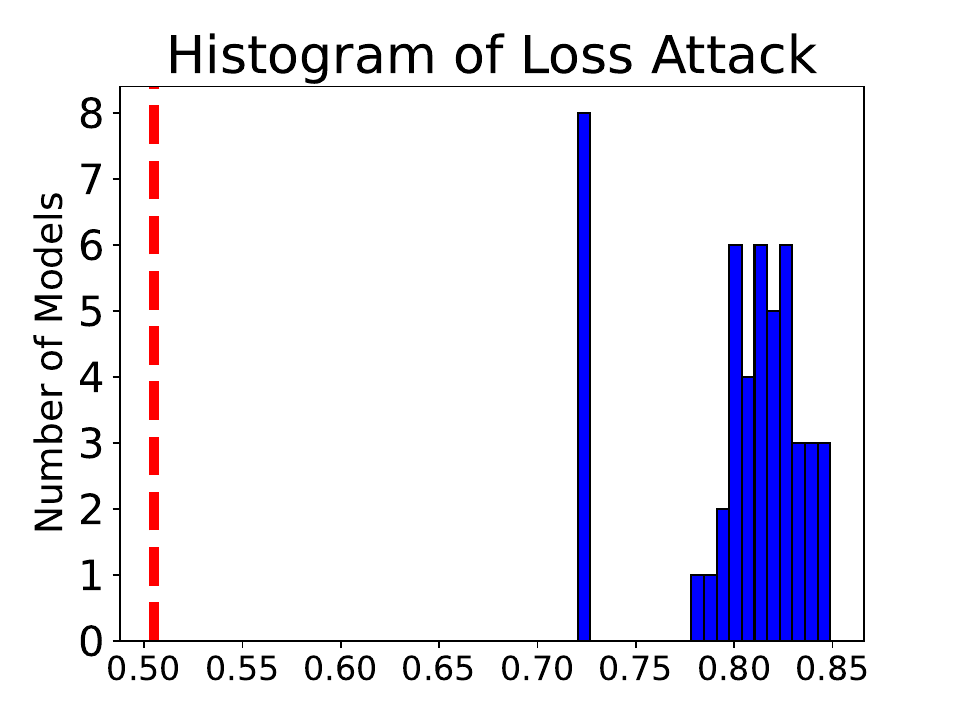}
    \includegraphics[width=0.495\linewidth]{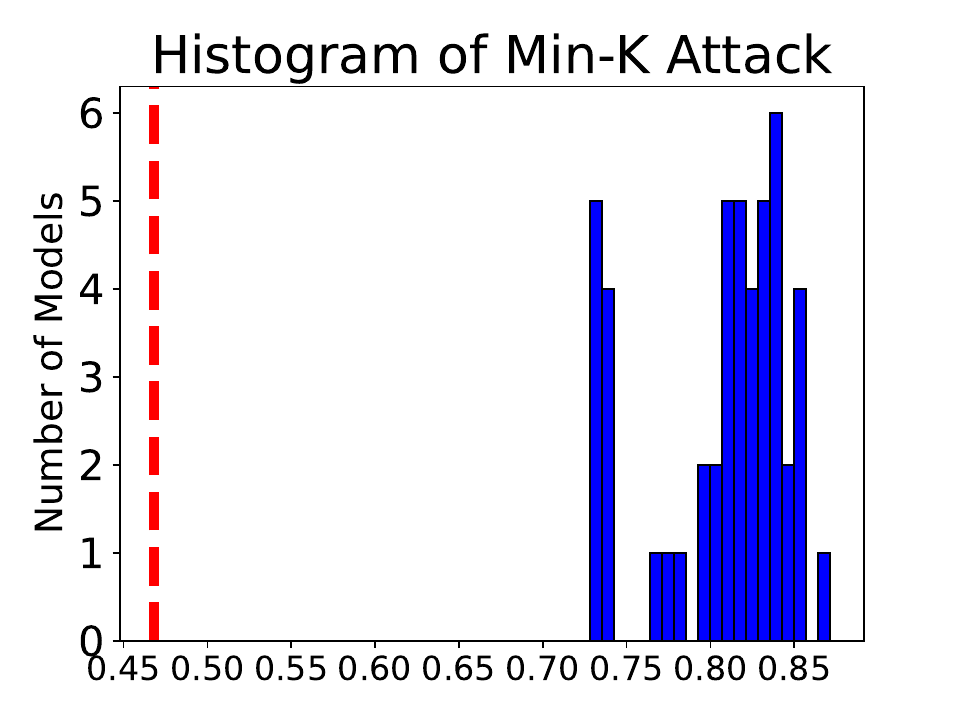}
    \includegraphics[width=0.495\linewidth]{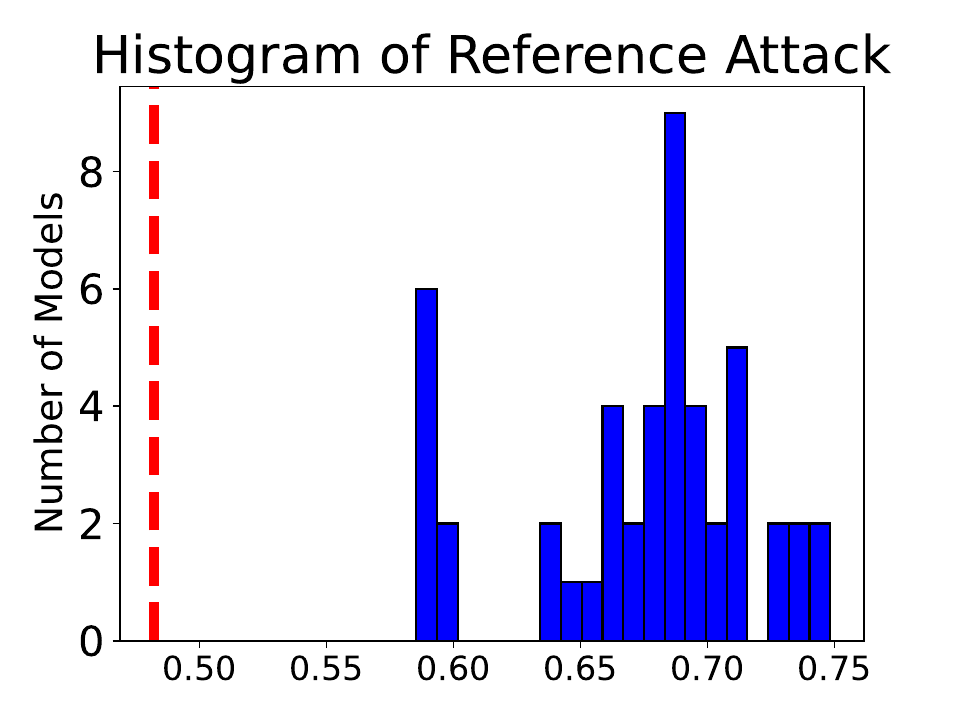}
    \includegraphics[width=0.495\linewidth]{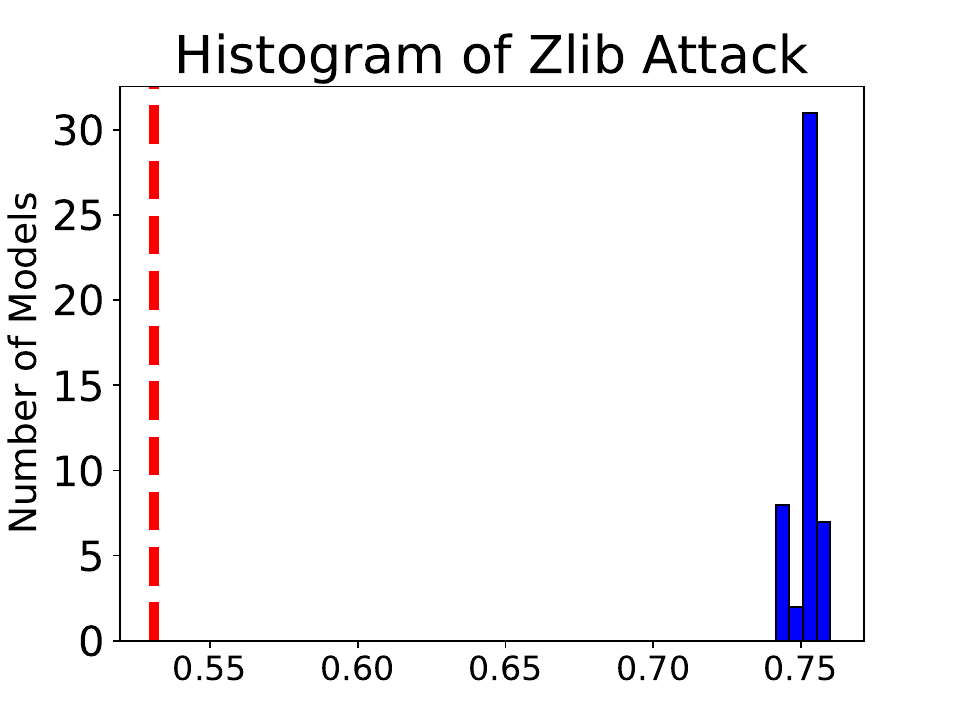}
    \caption{The distributions of the ROC-AUC score for MIA on models tuned with all the hyperparameter settings stated in Section~\ref{self-instruct-other-detail}. The red line shows the ROC-AUC score of the base Pythia-6.9b model. Blue bars represent the number of fine-tuned models. The x-axis denotes the ROC-AUC score.}
    \label{fig:histogram-max-success}
\end{figure}

Apart from the self-instruct tuned model with the top privacy risks, we also report the distributions of the ROC-AUC score for models with all the hyperparameter settings. The results are shown in Figure~\ref{fig:histogram-max-success}. We demonstrate that the ROC-AUC scores for all the models are higher than the base models with a large margin. The worst improvement is still larger than the $20\%$ increment compared with the base model under different attacks. The results reflect that the self-instruct tuning may cause privacy risks in nearly all cases.

\subsection{Ablation Studies}
In this section, we explore the key factors that influence the models' privacy leakage after the self-instruct tuning. We explore the key factors stated in Section~\ref{self-instruct-other-detail}, including the temperature, learning rate, and datasets.

\par\noindent\textbf{The impact of temperature.}
Different temperature settings for the generator affect the quality of the input-output examples, leading to variations in their relevance and diversity. Therefore, we plot the averaged ROC-AUC score for different MIA methods on Pythia 6.9b, which is fine-tuned on the self-instruct data generated with different temperatures. As for other hyperparameters, we choose the learning rate to be $2\times 10^{-4}$ and the data size to be $250$. It is the same setting for the self-instruct tuned model with the highest privacy risk.

\begin{figure}[!h]
    \centering
    \includegraphics[width=0.6\linewidth]{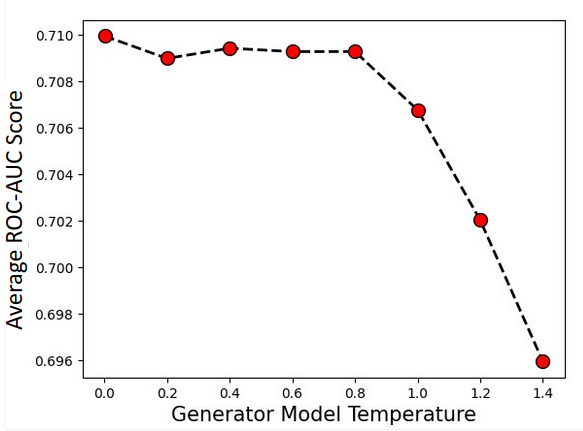}
    \caption{The averaged ROC-AUC score of different MIA methods conducted on self-instruct tuned models with different temperatures for the generator model.}
    \label{fig:self-instruct-mia-temp-averaged}
\end{figure}

Figure~\ref{fig:self-instruct-mia-temp-averaged} illustrates that temperature variations have minimal effect on the ROC-AUC score, with the highest score of 0.71 observed near a temperature of 0 and the lowest score of 0.69 at a temperature of 1.4. However, a consistent trend emerges where an increase in temperature is associated with a gradual decline in the averaged ROC-AUC score. This occurs because a higher temperature in the generator model reduces the similarity between the generated data and the original pre-training dataset. Thus, LLM's memories of the training data and the ROC-AUC score will be weaker and we recommend using a larger temperature to alleviate the privacy risks.



\par
\begin{figure}[!h]
    \centering
    \includegraphics[width=0.6\linewidth]{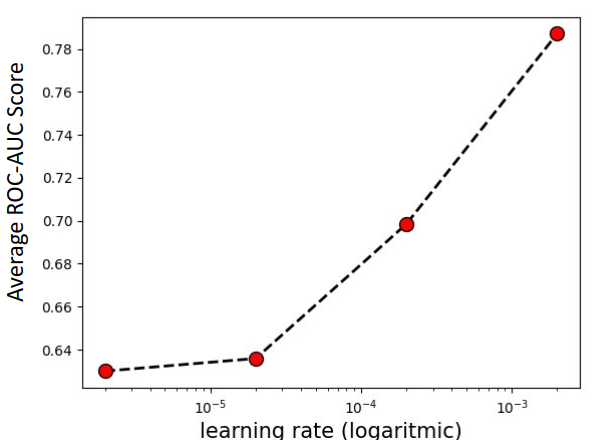}
    \caption{The averaged ROC-AUC score of different MIA methods conducted on self-instruct tuned models with various learning rates.}
    \label{fig:self-instruct-privacy-lr}
    \vspace{-1em}
\end{figure}

\noindent\textbf{The impact of learning rate.}
To investigate learning rate's effect in isolation, we fix the generator temperature to $0.6$ and dataset size to $250$. The results are drawn in Figure~\ref{fig:self-instruct-privacy-lr}. 
We observe a substantial correlation between ROC-AUC scores and increasing learning rates, with an improvement exceeding $20\%$ when the learning rate is adjusted from $2\times 10^{-6}$ to $2\times 10^{-3}$. A similar phenomenon is also observed when training with real data by previous studies~\cite{carlini2021extracting, carlini2023quantifyingmemorizationneurallanguage, huang2022arelargepretrainedlanguagemodelsleakingyourpersonalinformation}. A possible reason for such improvement is the larger learning rate makes LLMs better memorize the fine-tuning data and also activates the memorization of the pre-training datasets. Therefore, the MIA methods can perform better in such scenarios.

Apart from learning rates' influence on privacy, we also compare the model's utility fine-tuned with different learning rates. The results are listed in Table~\ref{tab:acclr}. Combined with the results in Figure~\ref{fig:self-instruct-privacy-lr}, we see that a larger learning rate will enhance both the utility and the privacy risks, as the models fit better in such settings. Moreover, we also find that using a smaller learning round $10^{-4}$ can reduce the AUC ROC's performance with a good performance.
\vspace{-0.5em}
\begin{table}[htbp] 
    \centering
    \small
    \phantom{xxx}
    
    \begin{tabular}{c|c c}
    \toprule
     & CaseHOLD & SSLA \\
    \midrule
    Learning Rate   & 7.8\% & 17.6\% \\
    $2\times 10^{-6}$ & 7.7\% & 17.3\% \\
    $2\times 10^{-5}$ & 8.2\% & 17.5\% \\
    $2\times 10^{-4}$ & 18.7\% & 18.5\% \\
    $2\times 10^{-3}$ & \textbf{22.0}\% & \textbf{23.0}\% \\
    \bottomrule
    \end{tabular}
    \caption{The accuracies for the pre-trained Pythia 6.9b and its self-instruct version with different learning rates.}
    \label{tab:acclr}
\end{table}


\par
\begin{figure}[!h]
    \centering
    \includegraphics[width=.495\linewidth]{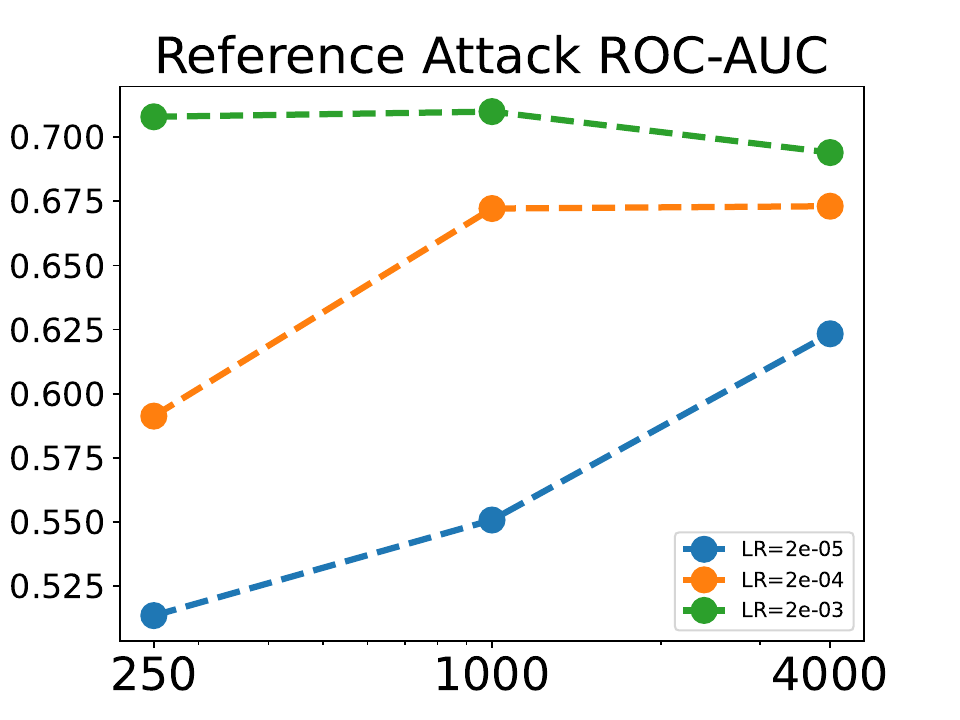}
    \includegraphics[width=.495\linewidth]{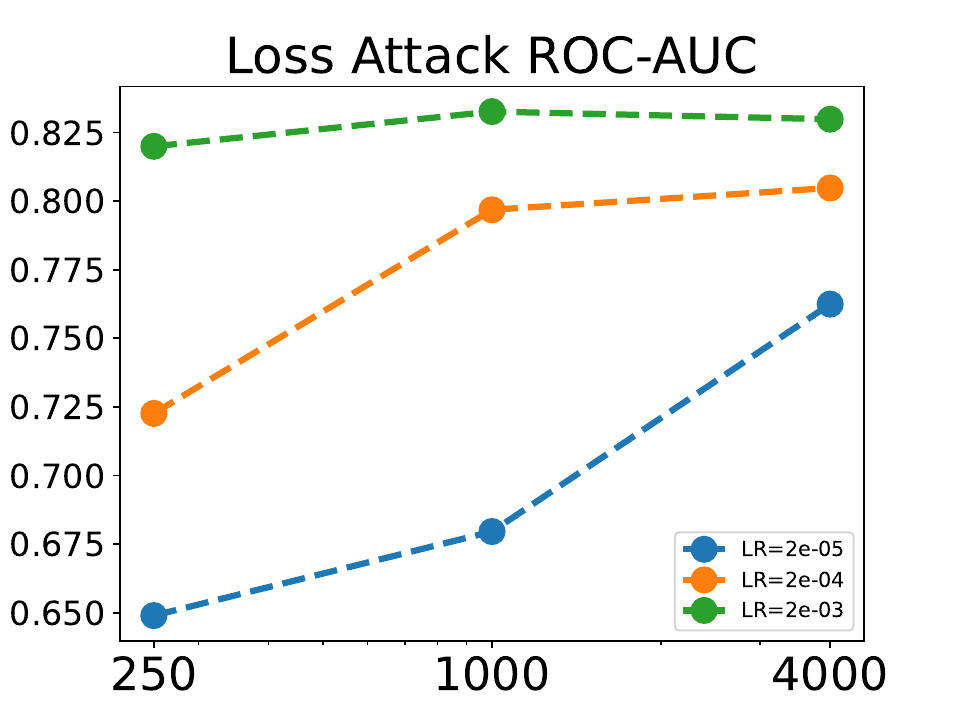}
    \includegraphics[width=.495\linewidth]{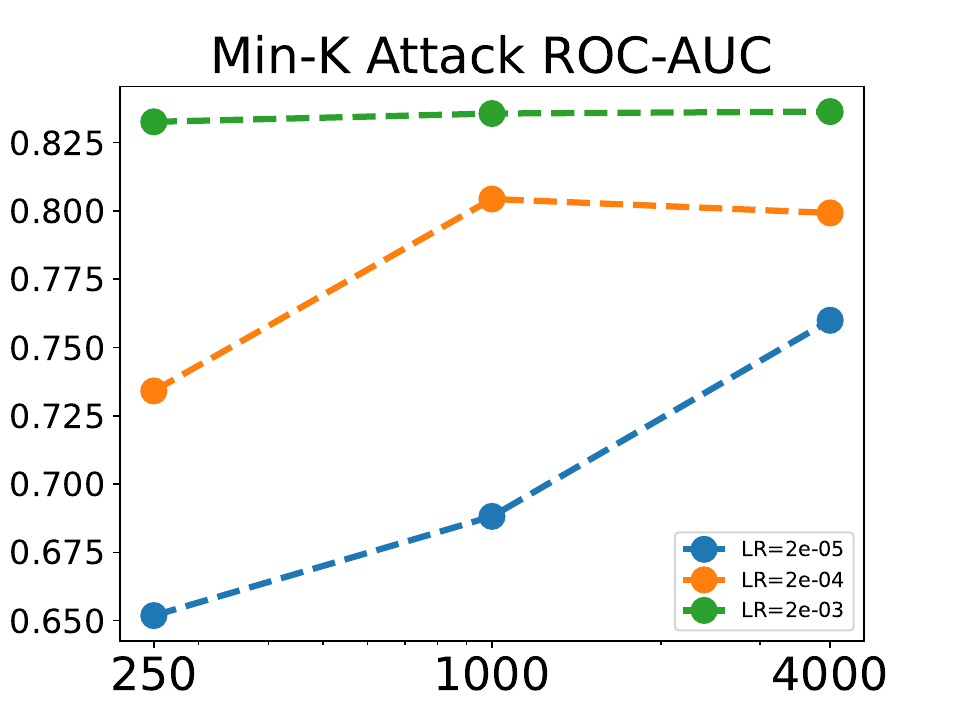}
    \includegraphics[width=.495\linewidth]{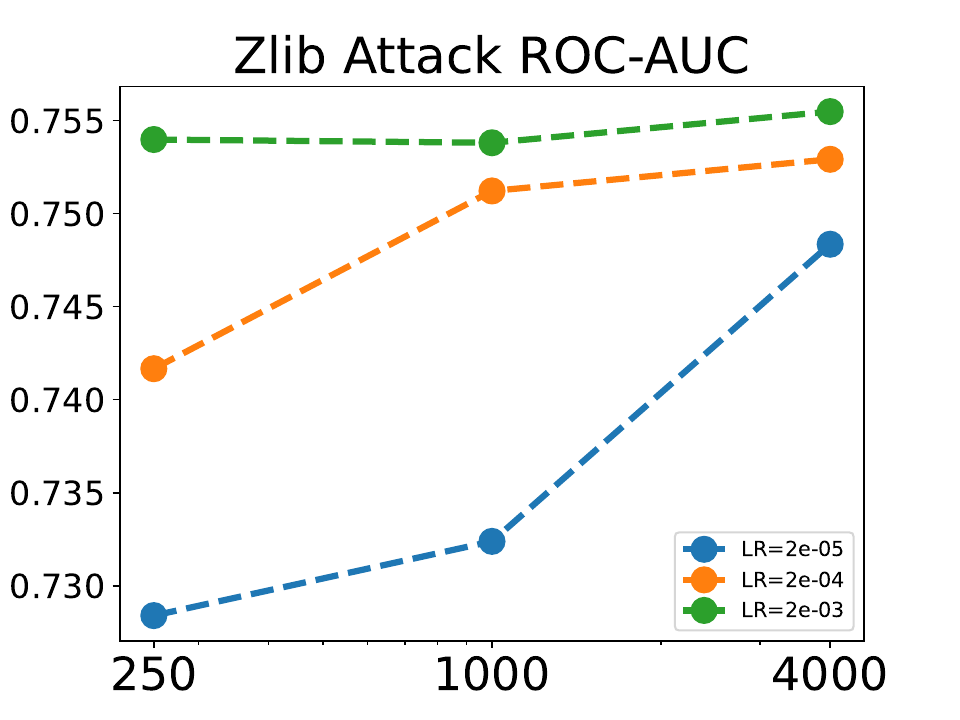}
   
    \caption{The ROC-AUC score of different MIA methods conducted on self-instruct tuned models with different data sizes. The x-axis denotes the data size while the y-axis denotes the ROC-AUC score.}
     \label{fig:self-instruct-dsize-5}
    \vspace{-1em}
\end{figure}

\noindent\textbf{The impact of dataset size.}
We explore how self-instruct dataset size impacts LLM privacy through experiments analyzing its effect on MIA performance. We plot the ROC-AUC scores for different MIA methods across models fine-tuned with datasets ranging from $250$ to $4,000$ samples, as shown in Figure~\ref{fig:self-instruct-dsize-5}. These experiments are conducted with three different learning rates: $2\times 10^{-5}$, $2\times 10^{-4}$, and $2\times 10^{-3}$.
The results indicate that all MIA methods display similar trends across different learning rates and datasets. A smaller learning rate notably enhances the ROC-AUC score as the dataset size increases. This is particularly evident for the reference-based attack, loss attack, and Min-k attack, where the ROC-AUC score improves by $10\%-20\%$ when the dataset size is scaled from $250$ to $4,000$. However, with larger learning rates, the differences between models fine-tuned with varying dataset sizes are less pronounced. This may be because models trained with smaller learning rates require more data to converge, while larger learning rates enable models to quickly memorize patterns similar to the original training samples in the self-instructed data, resulting in higher ROC-AUC scores after fine-tuning. Nevertheless, due to differences between the self-instructed data and the original pre-training data, the ROC-AUC score only increases to around 0.7-0.8. 
Overall, the findings suggest that both larger learning rates and increased dataset sizes can amplify privacy risks up to a certain threshold. Therefore, we recommend using a slightly smaller learning rate and dataset size to manage these risks effectively.

\subsection{Discussion}
In this section, we assess the privacy risks of LLMs on their pre-training datasets following self-instruct tuning. Using the example of a legal chatbot, we adopt the self-instruct pipeline to train a legal LLM based on Pythia 6.9b and evaluate both the model's utility and privacy. We observe that self-instruct tuning can substantially increase privacy risks for the pre-training dataset, FreeLaw, with over a $40\%$ improvement in ROC-AUC scores across various MIA methods. Additionally, our experiments reveal that the learning rate and dataset size are critical factors influencing privacy risks. Higher learning rates and larger datasets make the fine-tuned model more susceptible to membership inference attacks. Consequently, we recommend opting for a slightly lower learning rate and dataset size during training to safeguard privacy. While the suggested mitigations, such as using a lower learning rate (Section~\ref{Enron}) or higher temperatures (Section~\ref{Self-Instruct}), help to reduce the privacy risks, they do not entirely eliminate them, underscoring the need for further research and complementary approaches to fully address these vulnerabilities.

In addition, former methods such as differential privacy (DP)~\cite{li2022largelanguagemodelsstrongdifferentiallyprivatelearners, Behnia_2022privatelyfinetuningllmswithdifferentialprivacy}, data anonymization~\cite{gardiner-etal-2024-dataanonymization}, and data augmentation~\cite{ding2024dataaugmentationusingllms} can be used to alleviate the consequences of privacy leakage. DP can reduce memorization and mitigate membership inference, PII extraction, and model inversion risks by introducing noise~\cite{Behnia_2022privatelyfinetuningllmswithdifferentialprivacy}. Moreover, DP enables risk estimates through theoretical guarantees~\cite{Abadi_2016deeplearningwithdifferentialprivacy}. Data augmentation trains the model on multiple closely related data points rather than a single instance, potentially mitigating the impact of memorization~\cite{yu2021doesdataaugmentationaffectprivacy}. Yet, it is important to note that certain MIA implementations can leverage the structural similarities among augmented data points to enhance the success rate of attacks~\cite{li2024privacyeffectdataenhancementaugmentationincreasesmia}. Although data anonymization may not directly reduce memorization, it can render potential leakages less harmful and provide better safety standards, especially for PII attacks~\cite{gardiner-etal-2024-dataanonymization}. Future research should explore hybrid approaches that combine these techniques to enhance data privacy without significantly compromising model utility.

\section{Related Work}
\subsection{Privacy risks associated with LLMs}
Large Language Models (LLMs) have garnered significant attention due to their remarkable capabilities in natural language understanding. However, the rapid growth in model and dataset sizes has intensified concerns regarding privacy risks. Numerous studies~\cite{carlini2021extracting, carlini2023quantifyingmemorizationneurallanguage, duan2024membershipinferenceattackswork,huang2022arelargepretrainedlanguagemodelsleakingyourpersonalinformation, wang2024decodingtrustcomprehensiveassessmenttrustworthiness} have shown that larger and more sophisticated models are more vulnerable to pretraining data leakage and memorization, where data is inadvertently reproduced during generation.

\par This vulnerability has been rigorously quantified through methods such as Membership Inference Attacks (MIAs)~\cite{duan2024membershipinferenceattackswork, mireshghallah2022quantifyingprivacyrisksmaskedmia, carlini2021extracting}, which aim to determine whether a specific data point was used during the model's training, and Data Extraction Attacks \cite{carlini2021extracting}, which exploit the similarity between a target dataset and the model's output when prompted by an initial fragment of that data as an indicator of leakage.

\par  Prior work has shown open-source LLMs leak significant parts of their training data. Various methods such as data deduplication and differential privacy~\cite{li2022largelanguagemodelsstrongdifferentiallyprivatelearners} are proposed to mitigate the risks. 
However, these methods remain ineffective due to the computational infeasibility of implementing differentially private stochastic gradient descent. Additionally, evidence indicates that memorization can still compromise privacy, even in scenarios where observable overfitting is absent~\cite{tirumala2022memorization, carlini2021extracting}.

\par These findings underscore the urgent need for further research into more realistic scenarios, the practical effectiveness of proposed mitigation techniques, and their potential impact on model utility. There remains significant uncertainty about whether fine-tuning exacerbates or mitigates memorization~\cite{carlini2021extracting}, as well as the broader effects of different training settings on privacy risks. We draw attention to this uncertainty and fill the important gap of fine-tuning on generated data, which is crucial for the development of more secure and privacy-preserving LLMs.

\subsection{Privacy risks associated with synthetic  data}
Using synthetic data in deep learning has been a common practice for numerous purposes~\cite{ding2024dataaugmentationusingllms, yu2024finetuninglanguagemodelsgenerative}. A prominent use case of synthetic data is for training LLMs for downstream tasks~\cite{yu2024finetuninglanguagemodelsgenerative, wang2023selfinstructaligninglanguagemodels}. Recent works emphasized the efficacy of this use in terms of time and money\cite{wang2023selfinstructaligninglanguagemodels}. Some works, like self-play fine-tuning \cite{chen2024self}, also demonstrate that using self-generated synthetic data can further improve the model's performance. Furthermore, the possibility of using LLMs locally for data generation appeared to be a remedy for concerns about privacy in multi-party computing settings~\cite{syntheticclinicaltextmining}. In practice, more and more developers adopting synthetic data for their fine-tuning, e.g., Llama-3 \cite{llama3_license} and T\"ulu-3 \cite{lambert2024t}, current state-of-the-art LLMs, both adopt LLM-generated data in their training data.

\par However, the inherent risks of memorization and data leakage in LLMs raise concerns that fine-tuning on generated data may introduce significant, yet often overlooked privacy dangers. Specifically, generating data with a given prompt can lead to the reproduction of memorized data~\cite{wang2024decodingtrustcomprehensiveassessmenttrustworthiness}, a risk that parallels those seen in data extraction attacks~\cite{carlini2021extracting}. We note here that although has similar purposes, synthetic data is often created using statistical methods to mimic real data distribution. In contrast, LLM-generated data is produced from the model's already-learned representations, potentially memorized patterns from pretraining data. This makes LLM-generated data potentially more susceptible to privacy attacks, as it can inadvertently amplify the leakage of the original pretraining data. Finetuning on the memorized data can further exacerbate the privacy risks, by leaking PIIs from the pretraining corpus of the target, or the generator models.

In our work, we investigated the membership inference risks of LLMs fine-tuned on domain-specific generated data, which had not been addressed before. We demonstrate our findings in the highly sensitive domain of law with prominent open-source models for research purposes.

\section{Conclusion}
With the growing data requirements for fine-tuning, the use of generated data has become increasingly common. However, previous research has overlooked the potential privacy risks associated with fine-tuning models using generated data. In this paper, we address this gap by conducting experiments on two primary fine-tuning approaches with generated data: supervised fine-tuning with unstructured generated data and self-instruct tuning. We then evaluate the potential privacy risks involved in these fine-tuning pipelines. The results indicate LLMs can leak more private information on the related domain after fine-tuning with the generated data.

\section*{Acknowledgements}

We thank all anonymous reviewers for their constructive comments.
This work is partially funded by the European Health and Digital Executive Agency (HADEA) within the project ``Understanding the individual host response against Hepatitis D Virus to develop a personalized approach for the management of hepatitis D'' (DSolve, grant agreement number 101057917) and the BMBF with the project ``Repräsentative, synthetische Gesundheitsdaten mit starken Privatsphärengarantien'' (PriSyn, 16KISAO29K).








%
\newpage
\bibliographystyle{plain} 
\bibliography{sample} 

\begin{thebibliography}{10}

\bibitem{gpt4_blog}
\url{https://openai.com/research/gpt-4}.

\bibitem{llama3_license}
\url{https://llama.meta.com/llama3/license/}.

\bibitem{mistral_license}
\url{https://mistral.ai/terms/}.

\bibitem{CourtListener}
\url{https://www.courtlistener.com/}.

\bibitem{MetaLlama3}
{I}ntroducing {M}eta {L}lama 3: {T}he most capable openly available {L}{L}{M} to date --- ai.meta.com.
\newblock \url{https://ai.meta.com/blog/meta-llama-3/}, 2024.

\bibitem{huggingfaceSentencetransformersSentence}
sentence-transformers ({S}entence {T}ransformers) --- huggingface.co.
\newblock \url{https://huggingface.co/sentence-transformers}, 2024.

\bibitem{Abadi_2016deeplearningwithdifferentialprivacy}
Martin Abadi, Andy Chu, Ian Goodfellow, H.~Brendan McMahan, Ilya Mironov, Kunal Talwar, and Li~Zhang.
\newblock Deep learning with differential privacy.
\newblock In {\em ACM Conference on Computer and Communications Security (CCS)}, CCS’16. ACM, 2016.

\bibitem{Antonello2020SelectingIC}
Richard Antonello, Javier Turek, and Alexander~G. Huth.
\newblock Selecting informative contexts improves language model fine-tuning.
\newblock In {\em Annual Meeting of the Association for Computational Linguistics (ACL)}, 2020.

\bibitem{BJNACDDFGHJKKCEEHHHJKLNOABCMOMK22}
Yuntao Bai, Andy Jones, Kamal Ndousse, Amanda Askell, Anna Chen, Nova DasSarma, Dawn Drain, Stanislav Fort, Deep Ganguli, Tom Henighan, Nicholas Joseph, Saurav Kadavath, Jackson Kernion, Tom Conerly, Sheer El-Showk, Nelson Elhage, Zac Hatfield-Dodds, Danny Hernandez, Tristan Hume, Scott Johnston, Shauna Kravec, Liane Lovitt, Neel Nanda, Catherine Olsson, Dario Amodei, Tom Brown, Jack Clark, Sam McCandlish, Chris Olah, Ben Mann, and Jared Kaplan.
\newblock {Training a Helpful and Harmless Assistant with Reinforcement Learning from Human Feedback}.
\newblock {\em {arXiv preprint arXiv:2204.05862}}, 2022.

\bibitem{Behnia_2022privatelyfinetuningllmswithdifferentialprivacy}
Rouzbeh Behnia, Mohammadreza~Reza Ebrahimi, Jason Pacheco, and Balaji Padmanabhan.
\newblock Ew-tune: A framework for privately fine-tuning large language models with differential privacy.
\newblock In {\em 2022 IEEE International Conference on Data Mining Workshops (ICDMW)}, 2022.

\bibitem{biderman2023pythiasuiteanalyzinglarge}
Stella Biderman, Hailey Schoelkopf, Quentin Anthony, Herbie Bradley, Kyle O'Brien, Eric Hallahan, Mohammad~Aflah Khan, Shivanshu Purohit, USVSN~Sai Prashanth, Edward Raff, Aviya Skowron, Lintang Sutawika, and Oskar van~der Wal.
\newblock Pythia: A suite for analyzing large language models across training and scaling.
\newblock {\em arXiv preprint arXiv:2304.01373}, 2023.

\bibitem{bommasani2022opportunitiesrisksfoundationmodels}
Rishi Bommasani, Drew~A. Hudson, Ehsan Adeli, Russ Altman, Simran Arora, Sydney von Arx, Michael~S. Bernstein, Jeannette Bohg, Antoine Bosselut, Emma Brunskill, Erik Brynjolfsson, Shyamal Buch, Dallas Card, Rodrigo Castellon, Niladri Chatterji, Annie Chen, Kathleen Creel, Jared~Quincy Davis, Dora Demszky, Chris Donahue, Moussa Doumbouya, Esin Durmus, Stefano Ermon, John Etchemendy, Kawin Ethayarajh, Li~Fei-Fei, Chelsea Finn, Trevor Gale, Lauren Gillespie, Karan Goel, Noah Goodman, Shelby Grossman, Neel Guha, Tatsunori Hashimoto, Peter Henderson, John Hewitt, Daniel~E. Ho, Jenny Hong, Kyle Hsu, Jing Huang, Thomas Icard, Saahil Jain, Dan Jurafsky, Pratyusha Kalluri, Siddharth Karamcheti, Geoff Keeling, Fereshte Khani, Omar Khattab, Pang~Wei Koh, Mark Krass, Ranjay Krishna, Rohith Kuditipudi, Ananya Kumar, Faisal Ladhak, Mina Lee, Tony Lee, Jure Leskovec, Isabelle Levent, Xiang~Lisa Li, Xuechen Li, Tengyu Ma, Ali Malik, Christopher~D. Manning, Suvir Mirchandani, Eric Mitchell, Zanele Munyikwa, Suraj Nair,
  Avanika Narayan, Deepak Narayanan, Ben Newman, Allen Nie, Juan~Carlos Niebles, Hamed Nilforoshan, Julian Nyarko, Giray Ogut, Laurel Orr, Isabel Papadimitriou, Joon~Sung Park, Chris Piech, Eva Portelance, Christopher Potts, Aditi Raghunathan, Rob Reich, Hongyu Ren, Frieda Rong, Yusuf Roohani, Camilo Ruiz, Jack Ryan, Christopher Ré, Dorsa Sadigh, Shiori Sagawa, Keshav Santhanam, Andy Shih, Krishnan Srinivasan, Alex Tamkin, Rohan Taori, Armin~W. Thomas, Florian Tramèr, Rose~E. Wang, William Wang, Bohan Wu, Jiajun Wu, Yuhuai Wu, Sang~Michael Xie, Michihiro Yasunaga, Jiaxuan You, Matei Zaharia, Michael Zhang, Tianyi Zhang, Xikun Zhang, Yuhui Zhang, Lucia Zheng, Kaitlyn Zhou, and Percy Liang.
\newblock On the opportunities and risks of foundation models.
\newblock {\em arXiv preprint arXiv:2108.07258}, 2022.

\bibitem{BMRSKDNSSAAHKHCRZWWHCSLGCCBMRSA20}
Tom~B. Brown, Benjamin Mann, Nick Ryder, Melanie Subbiah, Jared Kaplan, Prafulla Dhariwal, Arvind Neelakantan, Pranav Shyam, Girish Sastry, Amanda Askell, Sandhini Agarwal, Ariel Herbert{-}Voss, Gretchen Krueger, Tom Henighan, Rewon Child, Aditya Ramesh, Daniel~M. Ziegler, Jeffrey Wu, Clemens Winter, Christopher Hesse, Mark Chen, Eric Sigler, Mateusz Litwin, Scott Gray, Benjamin Chess, Jack Clark, Christopher Berner, Sam McCandlish, Alec Radford, Ilya Sutskever, and Dario Amodei.
\newblock {Language Models are Few-Shot Learners}.
\newblock In {\em Advances in Neural Information Processing Systems (NeurIPS)}, 2020.

\bibitem{carlini2022membershipinferenceattacksprinciplesreferenceattack}
Nicholas Carlini, Steve Chien, Milad Nasr, Shuang Song, Andreas Terzis, and Florian Tramer.
\newblock Membership inference attacks from first principles.
\newblock {\em arXiv preprint arXiv:2112.03570}, 2022.

\bibitem{carlini2023quantifyingmemorizationneurallanguage}
Nicholas Carlini, Daphne Ippolito, Matthew Jagielski, Katherine Lee, Florian Tramer, and Chiyuan Zhang.
\newblock Quantifying memorization across neural language models.
\newblock In {\em International Conference on Learning Representations (ICLR)}, 2023.

\bibitem{carlini2021extracting}
Nicholas Carlini, Florian Tramer, Eric Wallace, Matthew Jagielski, Ariel Herbert-Voss, Katherine Lee, Adam Roberts, Tom Brown, Dawn Song, Ulfar Erlingsson, et~al.
\newblock Extracting training data from large language models.
\newblock In {\em USENIX Security Symposium (USENIX Security)}, 2021.

\bibitem{JanusInterface}
Xiaoyi Chen, Siyuan Tang, Rui Zhu, Shijun Yan, Lei Jin, Zihao Wang, Liya Su, Zhikun Zhang, XiaoFeng Wang, and Haixu Tang.
\newblock The janus interface: How fine-tuning in large language models amplifies the privacy risks.
\newblock {\em arXiv preprint arXiv:2310.15469}, 2024.

\bibitem{chen2024self}
Zixiang Chen, Yihe Deng, Huizhuo Yuan, Kaixuan Ji, and Quanquan Gu.
\newblock Self-play fine-tuning converts weak language models to strong language models.
\newblock {\em arXiv preprint arXiv:2401.01335}, 2024.

\bibitem{dettmers2023qloraefficientfinetuningquantized}
Tim Dettmers, Artidoro Pagnoni, Ari Holtzman, and Luke Zettlemoyer.
\newblock Qlora: Efficient finetuning of quantized llms.
\newblock {\em arXiv preprint arXiv:2305.14314}, 2023.

\bibitem{ding2024dataaugmentationusingllms}
Bosheng Ding, Chengwei Qin, Ruochen Zhao, Tianze Luo, Xinze Li, Guizhen Chen, Wenhan Xia, Junjie Hu, Anh~Tuan Luu, and Shafiq Joty.
\newblock Data augmentation using large language models: Data perspectives, learning paradigms and challenges.
\newblock {\em arXiv preprint arXiv:2403.02990}, 2024.

\bibitem{duan2024membershipinferenceattackswork}
Michael Duan, Anshuman Suri, Niloofar Mireshghallah, Sewon Min, Weijia Shi, Luke Zettlemoyer, Yulia Tsvetkov, Yejin Choi, David Evans, and Hannaneh Hajishirzi.
\newblock Do membership inference attacks work on large language models?
\newblock {\em arXiv preprint arXiv:2402.07841}, 2024.

\bibitem{duan2024uncoveringlatentmemoriesassessingmemorizationpatterns}
Sunny Duan, Mikail Khona, Abhiram Iyer, Rylan Schaeffer, and Ila~R Fiete.
\newblock Uncovering latent memories: Assessing data leakage and memorization patterns in frontier ai models.
\newblock {\em arXiv preprint arXiv:2406.14549}, 2024.

\bibitem{faker}
Daniele Faraglia.
\newblock Faker: Python package that generates fake data, 2025.

\bibitem{gao2020pile800gbdatasetdiverse}
Leo Gao, Stella Biderman, Sid Black, Laurence Golding, Travis Hoppe, Charles Foster, Jason Phang, Horace He, Anish Thite, Noa Nabeshima, Shawn Presser, and Connor Leahy.
\newblock The pile: An 800gb dataset of diverse text for language modeling.
\newblock {\em arXiv preprint arXiv:2101.00027}, 2020.

\bibitem{gardiner-etal-2024-dataanonymization}
Shayna Gardiner, Tania Habib, Kevin Humphreys, Masha Azizi, Frederic Mailhot, Anne Paling, Preston Thomas, and Nathan Zhang.
\newblock Data anonymization for privacy-preserving large language model fine-tuning on call transcripts.
\newblock In {\em Proceedings of the Workshop on Computational Approaches to Language Data Pseudonymization (CALD-pseudo)}, 2024.

\bibitem{legalbench}
Neel Guha, Julian Nyarko, Daniel~E. Ho, Christopher Ré, Adam Chilton, Aditya Narayana, Alex Chohlas-Wood, Austin Peters, Brandon Waldon, Daniel~N. Rockmore, Diego Zambrano, Dmitry Talisman, Enam Hoque, Faiz Surani, Frank Fagan, Galit Sarfaty, Gregory~M. Dickinson, Haggai Porat, Jason Hegland, Jessica Wu, Joe Nudell, Joel Niklaus, John Nay, Jonathan~H. Choi, Kevin Tobia, Margaret Hagan, Megan Ma, Michael Livermore, Nikon Rasumov-Rahe, Nils Holzenberger, Noam Kolt, Peter Henderson, Sean Rehaag, Sharad Goel, Shang Gao, Spencer Williams, Sunny Gandhi, Tom Zur, Varun Iyer, and Zehua Li.
\newblock Legalbench: A collaboratively built benchmark for measuring legal reasoning in large language models.
\newblock {\em arXiv preprint arXiv:2308.11462}, 2023.

\bibitem{hu2021loralowrankadaptationlarge}
Edward~J. Hu, Yelong Shen, Phillip Wallis, Zeyuan Allen-Zhu, Yuanzhi Li, Shean Wang, Lu~Wang, and Weizhu Chen.
\newblock Lora: Low-rank adaptation of large language models.
\newblock {\em arXiv preprint arXiv:2106.09685}, 2021.

\bibitem{huang2022arelargepretrainedlanguagemodelsleakingyourpersonalinformation}
Jie Huang, Hanyin Shao, and Kevin Chen-Chuan Chang.
\newblock Are large pre-trained language models leaking your personal information?
\newblock {\em arXiv preprint arXiv:2205.12628}, 2022.

\bibitem{jiang2024mora}
Ting Jiang, Shaohan Huang, Shengyue Luo, Zihan Zhang, Haizhen Huang, Furu Wei, Weiwei Deng, Feng Sun, Qi~Zhang, Deqing Wang, et~al.
\newblock Mora: High-rank updating for parameter-efficient fine-tuning.
\newblock {\em arXiv preprint arXiv:2405.12130}, 2024.

\bibitem{adam}
Diederik~P. Kingma and Jimmy Ba.
\newblock Adam: A method for stochastic optimization.
\newblock {\em arXiv preprint arXiv:1412.6980}, 2017.

\bibitem{theenroncorpus}
Bryan Klimt and Yiming Yang.
\newblock The enron corpus: A new dataset for email classification research.
\newblock In {\em European conference on machine learning (ECML)}, 2004.

\bibitem{kopiczko2023vera}
Dawid~J Kopiczko, Tijmen Blankevoort, and Yuki~M Asano.
\newblock Vera: Vector-based random matrix adaptation.
\newblock {\em arXiv preprint arXiv:2310.11454}, 2023.

\bibitem{lambert2024t}
Nathan Lambert, Jacob Morrison, Valentina Pyatkin, Shengyi Huang, Hamish Ivison, Faeze Brahman, Lester James~V Miranda, Alisa Liu, Nouha Dziri, Shane Lyu, et~al.
\newblock T$\backslash$" ulu 3: Pushing frontiers in open language model post-training.
\newblock {\em arXiv preprint arXiv:2411.15124}, 2024.

\bibitem{li2025salora}
Mingjie Li, Wai~Man Si, Michael Backes, Yang Zhang, and Yisen Wang.
\newblock Salora: Safety-alignment preserved low-rank adaptation.
\newblock {\em arXiv preprint arXiv:2501.01765}, 2025.

\bibitem{li2024privacyeffectdataenhancementaugmentationincreasesmia}
Xiao Li, Qiongxiu Li, Zhanhao Hu, and Xiaolin Hu.
\newblock On the privacy effect of data enhancement via the lens of memorization.
\newblock {\em arXiv preprint arXiv:2208.08270}, 2024.

\bibitem{li2022largelanguagemodelsstrongdifferentiallyprivatelearners}
Xuechen Li, Florian Tramèr, Percy Liang, and Tatsunori Hashimoto.
\newblock Large language models can be strong differentially private learners.
\newblock {\em arXiv preprint arXiv:2110.05679}, 2022.

\bibitem{liu2024dora}
Shih-Yang Liu, Chien-Yi Wang, Hongxu Yin, Pavlo Molchanov, Yu-Chiang~Frank Wang, Kwang-Ting Cheng, and Min-Hung Chen.
\newblock Dora: Weight-decomposed low-rank adaptation.
\newblock {\em arXiv preprint arXiv:2402.09353}, 2024.

\bibitem{LSSTWB23}
Nils Lukas, Ahmed Salem, Robert Sim, Shruti Tople, Lukas Wutschitz, and Santiago~Zanella B{\'{e}}guelin.
\newblock {Analyzing Leakage of Personally Identifiable Information in Language Models}.
\newblock In {\em {IEEE Symposium on Security and Privacy (S\&P)}}, 2023.

\bibitem{mireshghallah2022quantifyingprivacyrisksmaskedmia}
Fatemehsadat Mireshghallah, Kartik Goyal, Archit Uniyal, Taylor Berg-Kirkpatrick, and Reza Shokri.
\newblock Quantifying privacy risks of masked language models using membership inference attacks.
\newblock {\em arXiv preprint arXiv:2203.03929}, 2022.

\bibitem{nasr2023scalable}
Milad Nasr, Nicholas Carlini, Jonathan Hayase, Matthew Jagielski, A.~Feder Cooper, Daphne Ippolito, Christopher~A. Choquette{-}Choo, Eric Wallace, Florian Tram{\`{e}}r, and Katherine Lee.
\newblock {Scalable Extraction of Training Data from (Production) Language Models}.
\newblock {\em {arXiv preprint arXiv:2311.17035}}, 2023.

\bibitem{pickhardt2014generalizedlanguagemodelcombination}
Rene Pickhardt, Thomas Gottron, Martin Körner, Paul~Georg Wagner, Till Speicher, and Steffen Staab.
\newblock A generalized language model as the combination of skipped n-grams and modified kneser-ney smoothing.
\newblock {\em arXiv preprint arXiv:1404.3377}, 2014.

\bibitem{radford2018improving}
Alec Radford.
\newblock Improving language understanding by generative pre-training.
\newblock {\em OpenAI}, 2018.

\bibitem{radford2019language}
Alec Radford, Jeffrey Wu, Rewon Child, David Luan, Dario Amodei, Ilya Sutskever, et~al.
\newblock Language models are unsupervised multitask learners.
\newblock {\em OpenAI blog}, 2019.

\bibitem{seatgeek_thefuzz}
SeatGeek.
\newblock Thefuzz: Fuzzy string matching in python.
\newblock \url{https://github.com/seatgeek/thefuzz}, 2021.

\bibitem{SIHS21}
Virat Shejwalkar, Huseyin~A Inan, Amir Houmansadr, and Robert Sim.
\newblock {Membership Inference Attacks Against NLP Classification Models}.
\newblock In {\em Advances in Neural Information Processing Systems PriML Workshop (NeurIPS-PriML)}, 2021.

\bibitem{shi2024detectingpretrainingdatalargeminkattack}
Weijia Shi, Anirudh Ajith, Mengzhou Xia, Yangsibo Huang, Daogao Liu, Terra Blevins, Danqi Chen, and Luke Zettlemoyer.
\newblock {Detecting Pretraining Data from Large Language Models}.
\newblock {\em {arXiv preprint arXiv:2310.16789}}, 2023.

\bibitem{syntheticclinicaltextmining}
Ruixiang Tang, Xiaotian Han, Xiaoqian Jiang, and Xia Hu.
\newblock Does synthetic data generation of llms help clinical text mining?
\newblock {\em arXiv preprint arXiv:2303.04360}, 2023.

\bibitem{alpaca}
Rohan Taori, Ishaan Gulrajani, Tianyi Zhang, Yann Dubois, Xuechen Li, Carlos Guestrin, Percy Liang, and Tatsunori~B. Hashimoto.
\newblock Stanford alpaca: An instruction-following llama model, 2023.

\bibitem{tirumala2022memorization}
Kushal Tirumala, Aram Markosyan, Luke Zettlemoyer, and Armen Aghajanyan.
\newblock Memorization without overfitting: Analyzing the training dynamics of large language models.
\newblock In {\em Advances in Neural Information Processing Systems (NeurIPS)}, 2022.

\bibitem{touvron2023llama}
Hugo Touvron, Thibaut Lavril, Gautier Izacard, Xavier Martinet, Marie-Anne Lachaux, Timoth{\'e}e Lacroix, Baptiste Rozi{\`e}re, Naman Goyal, Eric Hambro, Faisal Azhar, et~al.
\newblock Llama: Open and efficient foundation language models.
\newblock {\em arXiv preprint arXiv:2302.13971}, 2023.

\bibitem{wang2024decodingtrustcomprehensiveassessmenttrustworthiness}
Boxin Wang, Weixin Chen, Hengzhi Pei, Chulin Xie, Mintong Kang, Chenhui Zhang, Chejian Xu, Zidi Xiong, Ritik Dutta, Rylan Schaeffer, Sang~T. Truong, Simran Arora, Mantas Mazeika, Dan Hendrycks, Zinan Lin, Yu~Cheng, Sanmi Koyejo, Dawn Song, and Bo~Li.
\newblock {DecodingTrust: {A} Comprehensive Assessment of Trustworthiness in {GPT} Models}.
\newblock {\em {arXiv preprint arXiv:2306.11698}}, 2023.

\bibitem{wang2023selfinstructaligninglanguagemodels}
Yizhong Wang, Yeganeh Kordi, Swaroop Mishra, Alisa Liu, Noah~A. Smith, Daniel Khashabi, and Hannaneh Hajishirzi.
\newblock Self-instruct: Aligning language models with self-generated instructions.
\newblock {\em arXiv preprint arXiv:2212.10560}, 2023.

\bibitem{yeom2018privacyriskmachinelearninglossbasedattack}
Samuel Yeom, Irene Giacomelli, Matt Fredrikson, and Somesh Jha.
\newblock Privacy risk in machine learning: Analyzing the connection to overfitting.
\newblock {\em arXiv preprint arXiv:1709.01604}, 2018.

\bibitem{yu2021doesdataaugmentationaffectprivacy}
Da~Yu, Huishuai Zhang, Wei Chen, Jian Yin, and Tie-Yan Liu.
\newblock How does data augmentation affect privacy in machine learning?
\newblock {\em arXiv preprint arXiv:2007.10567}, 2021.

\bibitem{yu2024finetuninglanguagemodelsgenerative}
Zhang~Ze Yu, Lau~Jia Jaw, Zhang Hui, and Bryan Kian~Hsiang Low.
\newblock Fine-tuning language models with generative adversarial reward modelling.
\newblock {\em arXiv preprint arXiv:2305.06176}, 2024.

\bibitem{zhang2022optopenpretrainedtransformerFacebookOPT}
Susan Zhang, Stephen Roller, Naman Goyal, Mikel Artetxe, Moya Chen, Shuohui Chen, Christopher Dewan, Mona Diab, Xian Li, Xi~Victoria Lin, Todor Mihaylov, Myle Ott, Sam Shleifer, Kurt Shuster, Daniel Simig, Punit~Singh Koura, Anjali Sridhar, Tianlu Wang, and Luke Zettlemoyer.
\newblock Opt: Open pre-trained transformer language models.
\newblock {\em arXiv preprint arXiv:2205.01068}, 2022.

\bibitem{zheng2021doespretraininghelpassessingcasehold}
Lucia Zheng, Neel Guha, Brandon~R. Anderson, Peter Henderson, and Daniel~E. Ho.
\newblock When does pretraining help? assessing self-supervised learning for law and the casehold dataset.
\newblock {\em arXiv preprint arXiv:2104.08671}, 2021.

\end{thebibliography}


@article{JanusInterface,
      title={The Janus Interface: How Fine-Tuning in Large Language Models Amplifies the Privacy Risks}, 
      author={Xiaoyi Chen and Siyuan Tang and Rui Zhu and Shijun Yan and Lei Jin and Zihao Wang and Liya Su and Zhikun Zhang and XiaoFeng Wang and Haixu Tang},
      year={2024},
      journal={arXiv preprint arXiv:2310.15469},
},
@article{radford2019language,
  title={Language models are unsupervised multitask learners},
  author={Radford, Alec and Wu, Jeffrey and Child, Rewon and Luan, David and Amodei, Dario and Sutskever, Ilya and others},
  journal={OpenAI blog},
  year={2019}
},
@article{radford2018improving,
  title={Improving language understanding by generative pre-training},
  author={Radford, Alec},
  journal={OpenAI},
  year={2018}
},

@inproceedings{shen2024donowcharacterizingevaluating,
      title={"Do Anything Now": Characterizing and Evaluating In-The-Wild Jailbreak Prompts on Large Language Models}, 
      author={Xinyue Shen and Zeyuan Chen and Michael Backes and Yun Shen and Yang Zhang},
      year={2024},
      booktitle={ACM Conference on Computer and Communications Security (CCS)}
},
@inproceedings{
wei2023jailbroken,
title={Jailbroken: How Does {LLM} Safety Training Fail?},
author={Alexander Wei and Nika Haghtalab and Jacob Steinhardt},
booktitle={Advance on Neural Information Processing Systems (NeurIPS)},
year={2023}
},
@inproceedings{Deng2023MASTERKEYAJ,
  title={MASTERKEY: Automated Jailbreaking of Large Language Model Chatbots},
  author={Gelei Deng and Yi Liu and Yuekang Li and Kailong Wang and Ying Zhang and Zefeng Li and Haoyu Wang and Tianwei Zhang and Yang Liu},
  booktitle={Proceedings 2024 Network and Distributed System Security Symposium (NDSS)},
  year={2023}
},
@article{
anonymous2024advprompter,
title={AdvPrompter: Fast Adaptive Adversarial Prompting for {LLM}s},
author={Anselm Paulus and Arman Zharmagambetov and Chuan Guo and Brandon Amos and Yuandong Tian},
journal={arXiv preprint arXiv:2404.16873},
year={2024}
},
@article{lambert2024t,
  title={T$\backslash$" ULU 3: Pushing Frontiers in Open Language Model Post-Training},
  author={Lambert, Nathan and Morrison, Jacob and Pyatkin, Valentina and Huang, Shengyi and Ivison, Hamish and Brahman, Faeze and Miranda, Lester James V and Liu, Alisa and Dziri, Nouha and Lyu, Shane and others},
  journal={arXiv preprint arXiv:2411.15124},
  year={2024}
},

@inproceedings{BleuInvention,
   author={Kishore Papineni and Salim Roukos and Todd Ward and Wei Jing Zhu},
   booktitle={Proceedings of the Annual Meeting of the Association for Computational Linguistics (ACL)},
   title={BLEU: A method for automatic evaluation of machine translation},
   year={2002},
},

@article{LlmsGoodPrivacyProtectionLearners,
      title={Large Language Models Can Be Good Privacy Protection Learners}, 
      author={Yijia Xiao and Yiqiao Jin and Yushi Bai and Yue Wu and Xianjun Yang and Xiao Luo and Wenchao Yu and Xujiang Zhao and Yanchi Liu and Haifeng Chen and Wei Wang and Wei Cheng},
      year={2023},
      journal={arXiv preprint arXiv:2310.02469},
},
@article{wang2024decodingtrustcomprehensiveassessmenttrustworthiness,
author = {Boxin Wang and Weixin Chen and Hengzhi Pei and Chulin Xie and Mintong Kang and Chenhui Zhang and Chejian Xu and Zidi Xiong and Ritik Dutta and Rylan Schaeffer and Sang T. Truong and Simran Arora and Mantas Mazeika and Dan Hendrycks and Zinan Lin and Yu Cheng and Sanmi Koyejo and Dawn Song and Bo Li},
title = {{DecodingTrust: {A} Comprehensive Assessment of Trustworthiness in {GPT} Models}},
journal = {{arXiv preprint arXiv:2306.11698}},
year = {2023}
},
@misc{llama3_license,
howpublished = {\url{https://llama.meta.com/llama3/license/}},
},
@misc{mistral_license,
howpublished = {\url{https://mistral.ai/terms/}},
},
@inproceedings{SIHS21,
author = {Virat Shejwalkar and Huseyin A Inan and Amir Houmansadr and Robert Sim},
title = {{Membership Inference Attacks Against NLP Classification Models}},
booktitle = {Advances in Neural Information Processing Systems PriML Workshop (NeurIPS-PriML)},
year = {2021}
},
@inproceedings{LSSTWB23,
author = {Nils Lukas and Ahmed Salem and Robert Sim and Shruti Tople and Lukas Wutschitz and Santiago Zanella B{\'{e}}guelin},
title = {{Analyzing Leakage of Personally Identifiable Information in Language Models}},
booktitle = {{IEEE Symposium on Security and Privacy (S\&P)}},
year = {2023}
},
@article{touvron2023llama,
  title={Llama: Open and efficient foundation language models},
  author={Touvron, Hugo and Lavril, Thibaut and Izacard, Gautier and Martinet, Xavier and Lachaux, Marie-Anne and Lacroix, Timoth{\'e}e and Rozi{\`e}re, Baptiste and Goyal, Naman and Hambro, Eric and Azhar, Faisal and others},
  journal={arXiv preprint arXiv:2302.13971},
  year={2023}
},
@inproceedings{carlini2021extracting,
  title={Extracting training data from large language models},
  author={Carlini, Nicholas and Tramer, Florian and Wallace, Eric and Jagielski, Matthew and Herbert-Voss, Ariel and Lee, Katherine and Roberts, Adam and Brown, Tom and Song, Dawn and Erlingsson, Ulfar and others},
  booktitle={USENIX Security Symposium (USENIX Security)},
  year={2021}
},
@article{chen2024self,
  title={Self-play fine-tuning converts weak language models to strong language models},
  author={Chen, Zixiang and Deng, Yihe and Yuan, Huizhuo and Ji, Kaixuan and Gu, Quanquan},
  journal={arXiv preprint arXiv:2401.01335},
  year={2024}
},
@article{zheng2021doespretraininghelpassessingcasehold,
      title={When Does Pretraining Help? Assessing Self-Supervised Learning for Law and the CaseHOLD Dataset}, 
      author={Lucia Zheng and Neel Guha and Brandon R. Anderson and Peter Henderson and Daniel E. Ho},
      year={2021},
      journal={arXiv preprint arXiv:2104.08671},
},
@misc{gpt4_blog,
howpublished = {\url{https://openai.com/research/gpt-4}},
},
@inproceedings{
privateandefficientLMwithsyntheticdata,
title={Training Private and Efficient Language Models with Synthetic Data from {LLM}s},
author={Da Yu and Arturs Backurs and Sivakanth Gopi and Huseyin Inan and Janardhan Kulkarni and Zinan Lin and Chulin Xie and Huishuai Zhang and Wanrong Zhang},
booktitle={Socially Responsible Language Modelling Research},
year={2023}
}, 

@article{syntheticclinicaltextmining,
      title={Does Synthetic Data Generation of LLMs Help Clinical Text Mining?}, 
      author={Ruixiang Tang and Xiaotian Han and Xiaoqian Jiang and Xia Hu},
      year={2023},
      journal={arXiv preprint arXiv:2303.04360},
},

@article{gao2020pile800gbdatasetdiverse,
      title={The Pile: An 800GB Dataset of Diverse Text for Language Modeling}, 
      author={Leo Gao and Stella Biderman and Sid Black and Laurence Golding and Travis Hoppe and Charles Foster and Jason Phang and Horace He and Anish Thite and Noa Nabeshima and Shawn Presser and Connor Leahy},
      year={2020},
      journal={arXiv preprint arXiv:2101.00027},
},

@article{biderman2023pythiasuiteanalyzinglarge,
      title={Pythia: A Suite for Analyzing Large Language Models Across Training and Scaling}, 
      author={Stella Biderman and Hailey Schoelkopf and Quentin Anthony and Herbie Bradley and Kyle O'Brien and Eric Hallahan and Mohammad Aflah Khan and Shivanshu Purohit and USVSN Sai Prashanth and Edward Raff and Aviya Skowron and Lintang Sutawika and Oskar van der Wal},
      year={2023},
      journal={arXiv preprint arXiv:2304.01373},
},

@article{ding2024dataaugmentationusingllms,
      title={Data Augmentation using Large Language Models: Data Perspectives, Learning Paradigms and Challenges}, 
      author={Bosheng Ding and Chengwei Qin and Ruochen Zhao and Tianze Luo and Xinze Li and Guizhen Chen and Wenhan Xia and Junjie Hu and Anh Tuan Luu and Shafiq Joty},
      year={2024},
      journal={arXiv preprint arXiv:2403.02990},
},

@article{yu2024finetuninglanguagemodelsgenerative,
      title={Fine-tuning Language Models with Generative Adversarial Reward Modelling}, 
      author={Zhang Ze Yu and Lau Jia Jaw and Zhang Hui and Bryan Kian Hsiang Low},
      year={2024},
      journal={arXiv preprint arXiv:2305.06176},
},

@article{perez2017effectivenessdataaugmentationimage,
      title={The Effectiveness of Data Augmentation in Image Classification using Deep Learning}, 
      author={Luis Perez and Jason Wang},
      year={2017},
      journal={arXiv preprint arXiv:1712.04621}
},

@article{wang2023selfinstructaligninglanguagemodels,
      title={Self-Instruct: Aligning Language Models with Self-Generated Instructions}, 
      author={Yizhong Wang and Yeganeh Kordi and Swaroop Mishra and Alisa Liu and Noah A. Smith and Daniel Khashabi and Hannaneh Hajishirzi},
      year={2023},
      journal={arXiv preprint arXiv:2212.10560},
},

@inproceedings{theenroncorpus,
  title={The enron corpus: A new dataset for email classification research},
  author={Klimt, Bryan and Yang, Yiming},
  booktitle={European conference on machine learning (ECML)},
  year={2004}
},

@article{perplexityinvention,
    author = {Jelinek, F. and Mercer, R. L. and Bahl, L. R. and Baker, J. K.},
    title = "{Perplexity—a measure of the difficulty of speech recognition tasks}",
    journal = {The Journal of the Acoustical Society of America},
    volume = {62},
    number = {S1},
    pages = {S63-S63},
    year = {2005}
},

@misc{seatgeek_thefuzz,
  author       = {SeatGeek},
  title        = {TheFuzz: Fuzzy String Matching in Python},
  year         = {2021},
  howpublished = {\url{https://github.com/seatgeek/thefuzz}}
},

@article{legalbench,
      title={LegalBench: A Collaboratively Built Benchmark for Measuring Legal Reasoning in Large Language Models}, 
      author={Neel Guha and Julian Nyarko and Daniel E. Ho and Christopher Ré and Adam Chilton and Aditya Narayana and Alex Chohlas-Wood and Austin Peters and Brandon Waldon and Daniel N. Rockmore and Diego Zambrano and Dmitry Talisman and Enam Hoque and Faiz Surani and Frank Fagan and Galit Sarfaty and Gregory M. Dickinson and Haggai Porat and Jason Hegland and Jessica Wu and Joe Nudell and Joel Niklaus and John Nay and Jonathan H. Choi and Kevin Tobia and Margaret Hagan and Megan Ma and Michael Livermore and Nikon Rasumov-Rahe and Nils Holzenberger and Noam Kolt and Peter Henderson and Sean Rehaag and Sharad Goel and Shang Gao and Spencer Williams and Sunny Gandhi and Tom Zur and Varun Iyer and Zehua Li},
      year={2023},
      journal={arXiv preprint arXiv:2308.11462},
},

@article{lawchatpaper,
      title={Adapting Large Language Models to Domains via Reading Comprehension}, 
      author={Daixuan Cheng and Shaohan Huang and Furu Wei},
      year={2024},
      journal={arXiv preprint arXiv:2309.09530},
},
@inproceedings{BMRSKDNSSAAHKHCRZWWHCSLGCCBMRSA20,
author = {Tom B. Brown and Benjamin Mann and Nick Ryder and Melanie Subbiah and Jared Kaplan and Prafulla Dhariwal and Arvind Neelakantan and Pranav Shyam and Girish Sastry and Amanda Askell and Sandhini Agarwal and Ariel Herbert{-}Voss and Gretchen Krueger and Tom Henighan and Rewon Child and Aditya Ramesh and Daniel M. Ziegler and Jeffrey Wu and Clemens Winter and Christopher Hesse and Mark Chen and Eric Sigler and Mateusz Litwin and Scott Gray and Benjamin Chess and Jack Clark and Christopher Berner and Sam McCandlish and Alec Radford and Ilya Sutskever and Dario Amodei},
title = {{Language Models are Few-Shot Learners}},
booktitle = {Advances in Neural Information Processing Systems (NeurIPS)},
year = {2020}
},
@inproceedings{Antonello2020SelectingIC,
  title={Selecting Informative Contexts Improves Language Model Fine-tuning},
  author={Richard Antonello and Javier Turek and Alexander G. Huth},
  booktitle={Annual Meeting of the Association for Computational Linguistics (ACL)},
  year={2020}
},
@misc{Black2021GPTNeoLS,
  title={GPT-Neo: Large Scale Autoregressive Language Modeling with Mesh-Tensorflow},
  author={Sid Black and Leo Gao and Phil Wang and Connor Leahy and Stella Biderman},
  year={2021},
  url={https://api.semanticscholar.org/CorpusID:245758737}
},
@article{pickhardt2014generalizedlanguagemodelcombination,
      title={A Generalized Language Model as the Combination of Skipped n-grams and Modified Kneser-Ney Smoothing}, 
      author={Rene Pickhardt and Thomas Gottron and Martin Körner and Paul Georg Wagner and Till Speicher and Steffen Staab},
      year={2014},
      journal={arXiv preprint arXiv:1404.3377},
},
@misc{MetaLlama3,
	author = {},
	title = {{I}ntroducing {M}eta {L}lama 3: {T}he most capable openly available {L}{L}{M} to date --- ai.meta.com},
	howpublished = {\url{https://ai.meta.com/blog/meta-llama-3/}},
	year = {2024}
},

@article{duan2024membershipinferenceattackswork,
      title={Do Membership Inference Attacks Work on Large Language Models?}, 
      author={Michael Duan and Anshuman Suri and Niloofar Mireshghallah and Sewon Min and Weijia Shi and Luke Zettlemoyer and Yulia Tsvetkov and Yejin Choi and David Evans and Hannaneh Hajishirzi},
      year={2024},
      journal={arXiv preprint arXiv:2402.07841},
},

@misc{CourtListener, 
howpublished = {\url{https://www.courtlistener.com/}},
} ,

@article{dettmers2023qloraefficientfinetuningquantized,
      title={QLoRA: Efficient Finetuning of Quantized LLMs}, 
      author={Tim Dettmers and Artidoro Pagnoni and Ari Holtzman and Luke Zettlemoyer},
      year={2023},
      journal={arXiv preprint arXiv:2305.14314}
},

@article{hu2021loralowrankadaptationlarge,
      title={LoRA: Low-Rank Adaptation of Large Language Models}, 
      author={Edward J. Hu and Yelong Shen and Phillip Wallis and Zeyuan Allen-Zhu and Yuanzhi Li and Shean Wang and Lu Wang and Weizhu Chen},
      year={2021},
      journal={arXiv preprint arXiv:2106.09685}
}


@inproceedings{tirumala2022memorization,
  title={Memorization without overfitting: Analyzing the training dynamics of large language models},
  author={Tirumala, Kushal and Markosyan, Aram and Zettlemoyer, Luke and Aghajanyan, Armen},
  booktitle={Advances in Neural Information Processing Systems (NeurIPS)},
  year={2022}
},
@article{yenduri2023generativepretrainedtransformercomprehensive,
      title={Generative Pre-trained Transformer: A Comprehensive Review on Enabling Technologies, Potential Applications, Emerging Challenges, and Future Directions}, 
      author={Gokul Yenduri and Ramalingam M and Chemmalar Selvi G and Supriya Y and Gautam Srivastava and Praveen Kumar Reddy Maddikunta and Deepti Raj G and Rutvij H Jhaveri and Prabadevi B and Weizheng Wang and Athanasios V. Vasilakos and Thippa Reddy Gadekallu},
      year={2023},
      journal={arXiv preprint arXiv:2305.10435}
},
@article{huang2022arelargepretrainedlanguagemodelsleakingyourpersonalinformation,
      title={Are Large Pre-Trained Language Models Leaking Your Personal Information?}, 
      author={Jie Huang and Hanyin Shao and Kevin Chen-Chuan Chang},
      year={2022},
      journal={arXiv preprint arXiv:2205.12628},
},


@inproceedings{carlini2023quantifyingmemorizationneurallanguage,
      title={Quantifying Memorization Across Neural Language Models}, 
      author={Nicholas Carlini and Daphne Ippolito and Matthew Jagielski and Katherine Lee and Florian Tramer and Chiyuan Zhang},
      year={2023},
      booktitle={International Conference on Learning Representations (ICLR)},
},

@article{fu2024practicalmembershipinferenceattacksselfcalibration,
      title={Practical Membership Inference Attacks against Fine-tuned Large Language Models via Self-prompt Calibration}, 
      author={Wenjie Fu and Huandong Wang and Chen Gao and Guanghua Liu and Yong Li and Tao Jiang},
      year={2024},
      journal={arXiv preprint arXiv:2311.06062},
},

@article{mireshghallah2022quantifyingprivacyrisksmaskedmia,
      title={Quantifying Privacy Risks of Masked Language Models Using Membership Inference Attacks}, 
      author={Fatemehsadat Mireshghallah and Kartik Goyal and Archit Uniyal and Taylor Berg-Kirkpatrick and Reza Shokri},
      year={2022},
      journal={arXiv preprint arXiv:2203.03929}
},

@article{bommasani2022opportunitiesrisksfoundationmodels,
      title={On the Opportunities and Risks of Foundation Models}, 
      author={Rishi Bommasani and Drew A. Hudson and Ehsan Adeli and Russ Altman and Simran Arora and Sydney von Arx and Michael S. Bernstein and Jeannette Bohg and Antoine Bosselut and Emma Brunskill and Erik Brynjolfsson and Shyamal Buch and Dallas Card and Rodrigo Castellon and Niladri Chatterji and Annie Chen and Kathleen Creel and Jared Quincy Davis and Dora Demszky and Chris Donahue and Moussa Doumbouya and Esin Durmus and Stefano Ermon and John Etchemendy and Kawin Ethayarajh and Li Fei-Fei and Chelsea Finn and Trevor Gale and Lauren Gillespie and Karan Goel and Noah Goodman and Shelby Grossman and Neel Guha and Tatsunori Hashimoto and Peter Henderson and John Hewitt and Daniel E. Ho and Jenny Hong and Kyle Hsu and Jing Huang and Thomas Icard and Saahil Jain and Dan Jurafsky and Pratyusha Kalluri and Siddharth Karamcheti and Geoff Keeling and Fereshte Khani and Omar Khattab and Pang Wei Koh and Mark Krass and Ranjay Krishna and Rohith Kuditipudi and Ananya Kumar and Faisal Ladhak and Mina Lee and Tony Lee and Jure Leskovec and Isabelle Levent and Xiang Lisa Li and Xuechen Li and Tengyu Ma and Ali Malik and Christopher D. Manning and Suvir Mirchandani and Eric Mitchell and Zanele Munyikwa and Suraj Nair and Avanika Narayan and Deepak Narayanan and Ben Newman and Allen Nie and Juan Carlos Niebles and Hamed Nilforoshan and Julian Nyarko and Giray Ogut and Laurel Orr and Isabel Papadimitriou and Joon Sung Park and Chris Piech and Eva Portelance and Christopher Potts and Aditi Raghunathan and Rob Reich and Hongyu Ren and Frieda Rong and Yusuf Roohani and Camilo Ruiz and Jack Ryan and Christopher Ré and Dorsa Sadigh and Shiori Sagawa and Keshav Santhanam and Andy Shih and Krishnan Srinivasan and Alex Tamkin and Rohan Taori and Armin W. Thomas and Florian Tramèr and Rose E. Wang and William Wang and Bohan Wu and Jiajun Wu and Yuhuai Wu and Sang Michael Xie and Michihiro Yasunaga and Jiaxuan You and Matei Zaharia and Michael Zhang and Tianyi Zhang and Xikun Zhang and Yuhui Zhang and Lucia Zheng and Kaitlyn Zhou and Percy Liang},
      year={2022},
      journal={arXiv preprint arXiv:2108.07258}
}

@article{li2022largelanguagemodelsstrongdifferentiallyprivatelearners,
      title={Large Language Models Can Be Strong Differentially Private Learners}, 
      author={Xuechen Li and Florian Tramèr and Percy Liang and Tatsunori Hashimoto},
      year={2022},
      journal={arXiv preprint arXiv:2110.05679},
},

@article{adam,
      title={Adam: A Method for Stochastic Optimization}, 
      author={Diederik P. Kingma and Jimmy Ba},
      year={2017},
      journal={arXiv preprint arXiv:1412.6980}
},

@article{shokri2017membershipinferenceattacksmachine,
      title={Membership Inference Attacks against Machine Learning Models}, 
      author={Reza Shokri and Marco Stronati and Congzheng Song and Vitaly Shmatikov},
      year={2017},
      journal={arXiv preprint arXiv:1610.05820},
},

@article{yeom2018privacyriskmachinelearninglossbasedattack,
      title={Privacy Risk in Machine Learning: Analyzing the Connection to Overfitting}, 
      author={Samuel Yeom and Irene Giacomelli and Matt Fredrikson and Somesh Jha},
      year={2018},
      journal={arXiv preprint arXiv:1709.01604},
},

@article{carlini2022membershipinferenceattacksprinciplesreferenceattack,
      title={Membership Inference Attacks From First Principles}, 
      author={Nicholas Carlini and Steve Chien and Milad Nasr and Shuang Song and Andreas Terzis and Florian Tramer},
      year={2022},
      journal={arXiv preprint arXiv:2112.03570},
},

@misc{huggingfaceSentencetransformersSentence,
	author = {},
	title = {sentence-transformers ({S}entence {T}ransformers) --- huggingface.co},
	howpublished = {\url{https://huggingface.co/sentence-transformers}},
	year = {2024}
},

@inproceedings{Behnia_2022privatelyfinetuningllmswithdifferentialprivacy,
   title={EW-Tune: A Framework for Privately Fine-Tuning Large Language Models with Differential Privacy},
   booktitle={2022 IEEE International Conference on Data Mining Workshops (ICDMW)},
   author={Behnia, Rouzbeh and Ebrahimi, Mohammadreza Reza and Pacheco, Jason and Padmanabhan, Balaji},
   year={2022} }

@inproceedings{Abadi_2016deeplearningwithdifferentialprivacy, series={CCS’16},
   title={Deep Learning with Differential Privacy},
   booktitle={ACM Conference on Computer and Communications Security (CCS)},
   publisher={ACM},
   author={Abadi, Martin and Chu, Andy and Goodfellow, Ian and McMahan, H. Brendan and Mironov, Ilya and Talwar, Kunal and Zhang, Li},
   year={2016}}

@article{yu2021doesdataaugmentationaffectprivacy,
      title={How Does Data Augmentation Affect Privacy in Machine Learning?}, 
      author={Da Yu and Huishuai Zhang and Wei Chen and Jian Yin and Tie-Yan Liu},
      year={2021},
      journal={arXiv preprint arXiv:2007.10567}
}

@article{li2024privacyeffectdataenhancementaugmentationincreasesmia,
      title={On the Privacy Effect of Data Enhancement via the Lens of Memorization}, 
      author={Xiao Li and Qiongxiu Li and Zhanhao Hu and Xiaolin Hu},
      year={2024},
      journal={arXiv preprint arXiv:2208.08270}
}

@article{duan2024uncoveringlatentmemoriesassessingmemorizationpatterns,
      title={Uncovering Latent Memories: Assessing Data Leakage and Memorization Patterns in Frontier AI Models}, 
      author={Sunny Duan and Mikail Khona and Abhiram Iyer and Rylan Schaeffer and Ila R Fiete},
      year={2024},
      journal={arXiv preprint arXiv:2406.14549}
}

@article{kopiczko2023vera,
  title={Vera: Vector-based random matrix adaptation},
  author={Kopiczko, Dawid J and Blankevoort, Tijmen and Asano, Yuki M},
  journal={arXiv preprint arXiv:2310.11454},
  year={2023}
}

@inproceedings{gardiner-etal-2024-dataanonymization,
    title = {Data Anonymization for Privacy-Preserving Large Language Model Fine-Tuning on Call Transcripts},
    author = {Gardiner, Shayna  and
      Habib, Tania  and
      Humphreys, Kevin  and
      Azizi, Masha  and
      Mailhot, Frederic  and
      Paling, Anne  and
      Thomas, Preston  and
      Zhang, Nathan},
    booktitle = {Proceedings of the Workshop on Computational Approaches to Language Data Pseudonymization (CALD-pseudo)},
    year = {2024}
}

@article{shi2024detectingpretrainingdatalargeminkattack,
author = {Weijia Shi and Anirudh Ajith and Mengzhou Xia and Yangsibo Huang and Daogao Liu and Terra Blevins and Danqi Chen and Luke Zettlemoyer},
title = {{Detecting Pretraining Data from Large Language Models}},
journal = {{arXiv preprint arXiv:2310.16789}},
year = {2023}
},

@article{zhang2022optopenpretrainedtransformerFacebookOPT,
      title={OPT: Open Pre-trained Transformer Language Models}, 
      author={Susan Zhang and Stephen Roller and Naman Goyal and Mikel Artetxe and Moya Chen and Shuohui Chen and Christopher Dewan and Mona Diab and Xian Li and Xi Victoria Lin and Todor Mihaylov and Myle Ott and Sam Shleifer and Kurt Shuster and Daniel Simig and Punit Singh Koura and Anjali Sridhar and Tianlu Wang and Luke Zettlemoyer},
      year={2022},
      journal={arXiv preprint arXiv:2205.01068},
},
@article{BJNACDDFGHJKKCEEHHHJKLNOABCMOMK22,
author = {Yuntao Bai and Andy Jones and Kamal Ndousse and Amanda Askell and Anna Chen and Nova DasSarma and Dawn Drain and Stanislav Fort and Deep Ganguli and Tom Henighan and Nicholas Joseph and Saurav Kadavath and Jackson Kernion and Tom Conerly and Sheer El-Showk and Nelson Elhage and Zac Hatfield-Dodds and Danny Hernandez and Tristan Hume and Scott Johnston and Shauna Kravec and Liane Lovitt and Neel Nanda and Catherine Olsson and Dario Amodei and Tom Brown and Jack Clark and Sam McCandlish and Chris Olah and Ben Mann and Jared Kaplan},
title = {{Training a Helpful and Harmless Assistant with Reinforcement Learning from Human Feedback}},
journal = {{arXiv preprint arXiv:2204.05862}},
year = {2022}
},
@article{nasr2023scalable,
author = {Milad Nasr and Nicholas Carlini and Jonathan Hayase and Matthew Jagielski and A. Feder Cooper and Daphne Ippolito and Christopher A. Choquette{-}Choo and Eric Wallace and Florian Tram{\`{e}}r and Katherine Lee},
title = {{Scalable Extraction of Training Data from (Production) Language Models}},
journal = {{arXiv preprint arXiv:2311.17035}},
year = {2023}
}

@misc{alpaca,
  author = {Rohan Taori and Ishaan Gulrajani and Tianyi Zhang and Yann Dubois and Xuechen Li and Carlos Guestrin and Percy Liang and Tatsunori B. Hashimoto },
  title = {Stanford Alpaca: An Instruction-following LLaMA model},
  year = {2023},
  publisher = {GitHub},
  journal = {GitHub repository}
}

@article{liu2024dora,
  title={Dora: Weight-decomposed low-rank adaptation},
  author={Liu, Shih-Yang and Wang, Chien-Yi and Yin, Hongxu and Molchanov, Pavlo and Wang, Yu-Chiang Frank and Cheng, Kwang-Ting and Chen, Min-Hung},
  journal={arXiv preprint arXiv:2402.09353},
  year={2024}
}

@misc{faker,
  author       = {Daniele Faraglia},
  title        = {Faker: Python package that generates fake data},
  year         = {2025},
  url          = {https://github.com/joke2k/faker}
}

@article{jiang2024mora,
  title={MoRA: High-Rank Updating for Parameter-Efficient Fine-Tuning},
  author={Jiang, Ting and Huang, Shaohan and Luo, Shengyue and Zhang, Zihan and Huang, Haizhen and Wei, Furu and Deng, Weiwei and Sun, Feng and Zhang, Qi and Wang, Deqing and others},
  journal={arXiv preprint arXiv:2405.12130},
  year={2024}
}
@article{li2025salora,
  title={SaLoRA: Safety-Alignment Preserved Low-Rank Adaptation},
  author={Li, Mingjie and Si, Wai Man and Backes, Michael and Zhang, Yang and Wang, Yisen},
  journal={arXiv preprint arXiv:2501.01765},
  year={2025}
}




\end{document}